\documentclass[aps,pre,twocolumn,showpacs,groupedaddress,
               showkeys,amsmath,amssymb]{revtex4}
\usepackage{graphicx}
\usepackage{bm}
\usepackage{bbm}
\usepackage{txfonts}
\usepackage{umlaut}

\usepackage{epic}
\usepackage{eepic}
\usepackage{pifont}

\usepackage{nicefrac}
\begin{document}

\title{How to model surface diffusion using the phase-field approach}

\author{Klaus Kassner} \affiliation{Institut f\"ur Theoretische
  Physik,
  Otto-von-Guericke-Universit\"at Magdeburg,\\
  Postfach 4120, D-39016 Magdeburg, Germany}

\date{\today}

\begin{abstract}
  It is demonstrated that the description of surface-diffusion
  controlled dynamics via the phase-field method is less trivial than
  it appears at first sight.  A seemingly straightforward approach
  previously used in the literature is shown to fail to produce the
  correct asymptotics, albeit in a subtle manner.  An apparently
  obvious alternative fails for a complementary reason.  Finally, a
  model is constructed that asymptotically approaches known
  sharp-interface equations without adding undesired constraints.  In
  order to provide a complete physical example, the model is exhibited
  for the elastically induced Grinfeld instability with material
  transport by surface diffusion.  The whole analysis is carried out
  for three-dimensional systems to pave the way for 
  simulations in %$\mathbb{R}^3$.
  $\varmathbb{R}^3$.

\end{abstract}

%46.25.-y Static elasticity
%47.20.Hw Morphological instability; phase changes

\maketitle


\newcommand{\s}{s}
\newcommand{\uc}{q}
\newcommand{\vel}{{\mathbf v}}
\newcommand{\Vel}{{\mathbf V}}
\newcommand{\gprime}{g^{\prime}}
\newcommand{\abs}[1]{\vert#1\vert}
\newcommand{\normhat}{\hat{{\mathbf n}}}
\newcommand{\tanghat}{\hat{{\mathbf t}}}   
\newcommand{\tangahat}{{\hat{\mathbf t}_1}}
\newcommand{\tangbhat}{{\hat{\mathbf t}_2}}
\newcommand{\chat}{\hat{c}}
\newcommand{\ghat}{\hat{G}}
\newcommand{\ez}{{\mathbf e}_z}
\newcommand{\er}{{\mathbf e}_r}
\newcommand{\es}{{\mathbf e}_s}
\newcommand{\eu}{{\mathbf e}_u}
\newcommand{\norm}{{\mathbf n}}
\newcommand{\tang}{{\mathbf t}}  
\newcommand{\tanga}{{\mathbf t}_1}
\newcommand{\tangb}{{\mathbf t}_2}
\newcommand{\rvec}{{\mathbf r}}
\newcommand{\Rvec}{{\mathbf R}}
\newcommand{\Uvec}{{\mathbf U}}
\newcommand{\wel}{{\mathbf v}}
\newcommand{\Wel}{{\mathbf V}}
\newcommand{\Avec}{{\mathbf A}}
\newcommand{\orderof}[1]{{\mathcal O}(#1)}
\newcommand{\linop}{{\mathcal L}}
\newcommand{\nablatwod}{{\nabla_{\rm 2D}}}
\newcommand{\lapltwod}{{\Delta_{\rm 2D}}}
\newcommand{\Er}{{\mathcal E}_r}
\newcommand{\Es}{{\mathcal E}_\s}
\newcommand{\Eu}{{\mathcal E}_\uc}
\newcommand{\Ei}{{\mathcal E}_i}
\newcommand{\Ej}{{\mathcal E}_j}
\newcommand{\Ealph}{{\mathcal E}_\alpha}
\newcommand{\Ebet}{{\mathcal E}_\beta}
\newcommand{\Ehr}{{\mathcal E}^r}
\newcommand{\Ehs}{{\mathcal E}^\s}
\newcommand{\Ehu}{{\mathcal E}^\uc}
\newcommand{\Ehi}{{\mathcal E}^i}
\newcommand{\Ehj}{{\mathcal E}^j}
\newcommand{\Ehalph}{{\mathcal E}^\alpha}
\newcommand{\Ehbet}{{\mathcal E}^\beta}
\newcommand{\Ehgam}{{\mathcal E}^\gamma}
\newcommand{\Ehmu}{{\mathcal E}^\mu}
\newcommand{\Ehnu}{{\mathcal E}^\nu}

\newcommand{\balpha}{{\bar{\alpha}}}
\newcommand{\bbeta}{{\bar{\beta}}}
\newcommand{\bgam}{{\bar{\gamma}}}
\newcommand{\bmu}{{\bar{\mu}}}
\newcommand{\bnu}{{\bar{\nu}}}

\newcommand{\Ehbalph}{{\mathcal E}^{\balpha}}
\newcommand{\Ehbbet}{{\mathcal E}^{\bbeta}}
\newcommand{\Ehbgam}{{\mathcal E}^{\bgam}}

\newcommand{\ssigma}{{\tilde{\sigma}}}
\newcommand{\Ssigma}{{\tilde{\Sigma}}}
\newcommand{\sigss}{\sigma_{\s\s}}
\newcommand{\siguu}{\sigma_{\uc\uc}}
\newcommand{\sigrr}{\sigma_{rr}}
\newcommand{\signn}{\sigma_{nn}}
\newcommand{\sigbalphbbet}{\sigma_{\balpha\bbeta}}
\newcommand{\sigbalphbalph}{\sigma_{\balpha\balpha}}

\newcommand{\phinul}{\phi^{(0)}}
\newcommand{\phione}{\phi^{(1)}}
\newcommand{\phitwo}{\phi^{(2)}}
\newcommand{\Phinul}{\Phi^{(0)}}
\newcommand{\Phione}{\Phi^{(1)}}
\newcommand{\Phitwo}{\Phi^{(2)}}

\newcommand{\phikalph}{\phi,_\alpha}
\newcommand{\phikbet}{\phi,_\beta}
\newcommand{\phikmu}{\phi,_\mu}
\newcommand{\Phikalph}{\Phi,_\alpha}
\newcommand{\Phikbet}{\Phi,_\beta}
\newcommand{\Phikmu}{\Phi,_\mu}
\newcommand{\Phikrho}{\Phi,_\rho}
\newcommand{\Phikrhonul}{\Phi,_\rho^{(0)}}
\newcommand{\Phikbalph}{\Phi,_\balpha}
\newcommand{\Phikbbet}{\Phi,_\bbeta}
\newcommand{\Phikbmu}{\Phi,_\bmu}
\newcommand{\Phikbnu}{\Phi,_\bnu}

\newcommand{\uijnul}{u_{ij}^{(0)}}
\newcommand{\uijone}{u_{ij}^{(1)}}
\newcommand{\uijtwo}{u_{ij}^{(2)}}
\newcommand{\Uijnul}{U_{ij}^{(0)}}
\newcommand{\Uijone}{U_{ij}^{(1)}}
\newcommand{\Uijtwo}{U_{ij}^{(2)}}
\newcommand{\Uabnul}{U_{\alpha\beta}^{(0)}}
\newcommand{\Uabone}{U_{\alpha\beta}^{(1)}}
\newcommand{\Uabtwo}{U_{\alpha\beta}^{(2)}}
\newcommand{\Unul}{U^{(0)}}
\newcommand{\Urrnul}{U_{rr}^{(0)}}
\newcommand{\Ussnul}{U_{\s\s}^{(0)}}
\newcommand{\Uuunul}{U_{\uc\uc}^{(0)}}
\newcommand{\Ursnul}{U_{r\s}^{(0)}}
\newcommand{\Urunul}{U_{r\uc}^{(0)}}
\newcommand{\Usunul}{U_{\s\uc}^{(0)}}
\newcommand{\Urr}{U_{rr}}
\newcommand{\Uss}{U_{\s\s}}
\newcommand{\Uuu}{U_{\uc\uc}}
\newcommand{\Urs}{U_{r\s}}
\newcommand{\Uru}{U_{r\uc}}
\newcommand{\Usu}{U_{\s\uc}}
\newcommand{\urrnul}{u_{rr}^{(0)}}
\newcommand{\ussnul}{u_{\s\s}^{(0)}}
\newcommand{\uuunul}{u_{\uc\uc}^{(0)}}
\newcommand{\urr}{u_{rr}}
\newcommand{\unn}{u_{nn}}
\newcommand{\uss}{u_{\s\s}}
\newcommand{\uuu}{u_{\uc\uc}}

\newcommand{\linopl}{{\mathcal L}}

\newcommand{\Gammagab}{\Gamma_{\alpha\beta}^{\gamma}}
\newcommand{\Gammagba}{\Gamma_{\beta\alpha}^{\gamma}}

\newcommand{\Gammarrr}{\Gamma_{rr}^{r}}
\newcommand{\Gammarss}{\Gamma_{\s\s}^{r}}
\newcommand{\Gammaruu}{\Gamma_{\uc\uc}^{r}}
\newcommand{\Gammarsr}{\Gamma_{\s r}^{r}}
\newcommand{\Gammarrs}{\Gamma_{r \s}^{r}}
\newcommand{\Gammarur}{\Gamma_{\uc r}^{r}}
\newcommand{\Gammarru}{\Gamma_{r \uc}^{r}}
\newcommand{\Gammarsu}{\Gamma_{\s\uc}^{r}}
\newcommand{\Gammarus}{\Gamma_{\uc\s}^{r}}

\newcommand{\Gammasss}{\Gamma_{\s\s}^{\s}}
\newcommand{\Gammasrr}{\Gamma_{rr}^{\s}}
\newcommand{\Gammasuu}{\Gamma_{\uc\uc}^{\s}}
\newcommand{\Gammassr}{\Gamma_{\s r}^{\s}}
\newcommand{\Gammasrs}{\Gamma_{r \s}^{\s}}
\newcommand{\Gammasur}{\Gamma_{\uc r}^{\s}}
\newcommand{\Gammasru}{\Gamma_{r \uc}^{\s}}
\newcommand{\Gammassu}{\Gamma_{\s\uc}^{\s}}
\newcommand{\Gammasus}{\Gamma_{\uc\s}^{\s}}

\newcommand{\Gammauuu}{\Gamma_{\uc\uc}^{\uc}}
\newcommand{\Gammaurr}{\Gamma_{rr}^{\uc}}
\newcommand{\Gammauss}{\Gamma_{\s\s}^{\uc}}
\newcommand{\Gammausr}{\Gamma_{\s r}^{\uc}}
\newcommand{\Gammaurs}{\Gamma_{r \s}^{\uc}}
\newcommand{\Gammauur}{\Gamma_{\uc r}^{\uc}}
\newcommand{\Gammauru}{\Gamma_{r \uc}^{\uc}}
\newcommand{\Gammausu}{\Gamma_{\s\uc}^{\uc}}
\newcommand{\Gammauus}{\Gamma_{\uc\s}^{\uc}}

\newcommand{\ud}{u}
\newcommand{\potelast}{V_{\rm el}}
\newcommand{\potelastnul}{V_{\rm el}^{(0)}}
\newcommand{\potelastnulbar}{\bar{V}_{\rm el}^{(0)}}
\newcommand{\lamtil}{\tilde{\lambda}}
\newcommand{\mus}{G}
\newcommand{\tnabla}{\tilde{\nabla}}
\newcommand{\hnul}{h_0}

%-----------------------------------------------

\renewcommand{\Vec}[1]{{\mathbf #1}}
\section{Introduction}
For a large class of pattern-forming systems, the essential dynamics to
be understood and described is that of an interface between two
phases.  Mathematically speaking, the problem to be solved consists in
determining the position of the interface as a function of time, i.e.,
it is a free or moving-boundary problem.

Phase-field models have been established as powerful tools for the
numerical simulation of this kind of problem.  They avoid explicit
front tracking and are versatile enough to deal with topological
changes.  In a phase-field model, information on the interface
position is present implicitly, given either as a level set of a
particular value of the phase field (in two-phase models) or by
equality of the phase-field values for different phases (in
multi-phase models), and can be recovered by computation of the
appropriate level set at only those times when knowledge of the
position is desired.

A major field of application are solidification problems, where
diffuse-interface models were developed early on
\cite{langer75,collins85,caginalp86} and have seen renewed interest
ever since computational power increased enough to render their
simulation feasible. The concept was extended to anisotropic interface
properties \cite{mcfadden93}, and first qualitative numerical
calculations of dendritic growth \cite{kobayashi93,kobayashi94} were
followed by theoretical improvement of the asymptotics permitting
quantitative simulations \cite{karma96,karma98}, at least for
intermediate to large undercoolings.  Non-dendritic growth
morphologies were also simulated, even in three dimensions
\cite{abel97}.  Generalizations included the description of the
coexistence of more than two phases \cite{wheeler96,garcke99}.

Additional examples of successful application of the tool phase field
include the modeling of step flow growth \cite{opl03,voigt04} and of
the elastically induced morphological instability
\cite{mueller99,kassner99,kassner01}, often labeled Grinfeld
\cite{grinfeld86} or Asaro-Tiller-Grinfeld (ATG) instability
\cite{asaro72}.  All of the examples mentioned so far dealt with {\em
  nonconservative} interface dynamics, where a particle reservoir is
provided by either the melt that is in contact with the solid or the
adatom phase on a vicinal surface.

Actually, regarding the ATG instability, which is an instability
with respect to material transport driven by elastic energy, interest
initially focused on transport by surface diffusion, which leads to
{\em conserved} dynamics.  This is the case in the first article by
Asaro and Tiller \cite{asaro72}, but also in the first numerical
simulations by sharp-interface continuum models \cite{yang93},
preceding computations of the instability under transport by
melting-crystallization \cite{kassner94}.

The situation reversed, when the phase-field method was for the first
time employed to compute the ATG instability
\cite{mueller99,kassner99}.  Here all the early works considered a
nonconserved phase-field
\cite{mueller99,kassner99,kassner01,haataja02}.  In fact, there seem
to be no publications so far treating conserved phase-field dynamics
in this system.  This difference in preferences when modeling either
on the basis of a sharp-interface model or using a phase field may be
due to the fact that writing down a nonconservative and a conservative
model is equally simple in the former case, whereas it is less
straightforward to write down the conservative model within the
phase-field approach -- correctly -- than the nonconservative one.
The demonstration of this assertion will be a major point to be made
in the present article.

This is not to say that phase-field models with a conservation law for
the phase field have not been considered at all.  Starting from a
Cahn-Hilliard equation with a concentration dependent mobility, Cahn
et al.~\cite{cahn96} obtained an interface equation with the normal
velocity proportional to the Laplacian of the mean curvature.  It then
looks as if all phase-field models with surface diffusion should be
derivable on the basis of similar considerations.  Indeed, comparable
models have been applied in the simulation of electromigration and voiding
in thin metal films \cite{mahadevan99a,bhate00}.  These two models are
slightly different, but the difference is not crucial and all
previous models seem to suffer from the same problem, to be discussed
in the following.

As we shall see, it is quite easy to set up a conservative phase-field
model. But it is more difficult to obtain the correct asymptotics
describing surface diffusion as given by the desired sharp-interface
limit. Past models such as the ones presented in
\cite{cahn96,mahadevan99a,bhate00} describe this asymptotics {\em
  almost} correctly, but not quite.  The purpose of this article is to
point out the reasons, to explore an alternative approach and finally,
to give a model that is asymptotically correct.

To be definite, I shall consider the case of surface diffusion in the
ATG instability.  The paper is organized as follows.  In
Sec.~\ref{sec:sharp_i_mod}, the sharp-interface model to be
approximated by the phase-field equations is specified.  Section
\ref{sec:scalar} then presents the standard approach that previously
was supposed to reduce to the correct limit and pinpoints the
oversight in hitherto existing asymptotic analyses.  An alternative
approach is presented in Sec.~\ref{sec:tensorial} failing for
complementary reasons.  By an appropriate combination of the ideas
from both approaches, I will then arrive at a phase-field model in
Sec.~\ref{sec:correct} that may lack elegance but has the virtue of
giving the correct asymptotic behaviour and doing so at a lower
effective expansion order than the preceding models.  All derivations
are given for the three-dimensional case.  While this is slightly more
involved than with the two-dimensional model, it is expected that
future interesting applications will be three-dimensional.
Phase-field models for conserved dynamics considered so far have been
restricted to two dimensions.  Moreover, the tensorial nature of the
elasticity part of the model makes a generalization from two to three
dimensions less straightforward than with scalar or vector equations.
Section \ref{sec:conclusions} contains some concluding remarks. A
number of mathematical details from differential geometry as well as
the asymptotics of the elastic part of the model, which is
unproblematic, are relegated to various appendices.

\section{Sharp-interface model for the ATG instability under
surface diffusion \label{sec:sharp_i_mod}}

One of the more convenient descriptions of the ATG instability starts
from an expression for the local chemical potential difference between
the solid and the second phase [liquid, gas (vapour), or vacuum] at
the interface.  In three dimensions, this is given by \cite{berger03}
\begin{eqnarray}
  \label{eq:delta_mu}
   \delta \mu  &=&  \frac{1}{2E \rho_s}
\bigg\{ (1+\nu)\sum_{i,j} \left(\sigma_{ij}-\sigma_{nn}\delta_{ij}\right)^2
\nonumber\\
&&
\mbox{} -\nu \Big[\sum_k \big(\sigma_{kk}-\sigma_{nn}\big)\Big]^2\bigg\}
\nonumber\\
&& \mbox{} +  \frac{1}{\rho_s} \, \gamma \left(\kappa_1+\kappa_2\right)
%+ \frac{\Delta W}{\rho_s}
\>.
\end{eqnarray}
In this equation, the material parameters are Young's modulus $E$, the
Poisson number $\nu$, meaning that we assume isotropic elastic
properties, the surface tension $\gamma$, assumed isotropic as well
for simplicity, and the density $\rho_s$ of the solid phase. The
$\sigma_{ij}$ are the (cartesian) components of the stress tensor,
$\sigma_{nn}$ is its normal component at the interface, $\kappa_1$ and
$\kappa_2$ are the two principal curvatures of the interface at the
point considered. Their sum is commonly denoted as mean curvature.  
% To compactify notation a bit, we will often abbreviate
% $\sigma^{-}_{ij}\equiv\sigma_{ij}-\sigma_{nn}\delta_{ij}$,
% $\kappa\equiv\kappa_1+\kappa_2$.
%and moreover use Einstein's summation convention,
%when no confusion can arise.
%The additive term $\Delta W$ is introduced to adjust the equilibrium
%state: if, for given environmental conditions (such as temperature or
%vapour pressure), the prestrained solid subject to a constant planar
%excess stress $\sigma_{00}$ is at (a possibly unstable) equilibrium,
%then
%\begin{equation}
%  \label{eq:delta_W}
%  \Delta W = - \frac{1}{2E} \bigg\{ (1+\nu)\sum_{i,j} \sigma_{00,ij}^2 -
%\nu\, \Big(\sum_k \sigma_{00,kk}\Big)^2\bigg\} \,,
%\end{equation}

In writing equation \eqref{eq:delta_mu}, we have neglected a
gravitational contribution to the chemical potential \cite{kassner01}
that usually is unimportant in situations, where surface diffusion
governs the dynamics. For example, in the self-organized formation of
patterns under strain built up during molecular beam epitaxy of
semiconductor heterostructures, the strain energy induced by lattice
mismatch exceeds any potential energy in the gravitational field by
several orders of magnitude.

To determine the stresses needed in the calculation of the chemical
potential, one has to solve the elastic equations.  These are given by
the mechanical equilibrium condition of vanishing divergence of the
stress tensor,
\begin{equation}
  \label{eq:mecheq}
  \sum_j \frac{\partial\sigma_{ij}}{\partial x_j} = 0 \>,
\end{equation}
where volume forces such as gravity are assumed to be absent and we
have restricted ourselves to the static limit of elasticity (hence
there are no sound waves),
% \footnote{We
%   have restricted ourselves to the static limit of elasticity, since
%   growth processes are slow on the time scale of sound propagation.},
plus a constitutive equation, for which linear elasticity, described by
Hooke's law
\begin{equation}
\sigma_{ij} = \frac{E}{1+\nu}\bigg(\ud_{ij}
+\frac{\nu}{1-2\nu} \sum_k \ud_{kk} \,\delta_{ij} \bigg) - p_0 \delta_{ij}
\>,
\label{eq:hooke}
\end{equation}
is assumed, in which
\begin{equation}
\ud_{ij}= \frac{1}{2}\left(\frac{\partial \ud_i}{\partial x_j} +
\frac{\partial \ud_j}{\partial x_i}\right)
\label{eq:strain}
\end{equation}
is the (small-)strain tensor.  It is convenient to choose the
hydrostatic strain state assumed by the solid under the equilibrium
pressure $p_0$ of the non-solid phase as reference state, hence the
term $- p_0 \delta_{ij}$ ($p_0$ may of course be zero).  This simply
means that we are measuring displacements of material points starting
from their position in this state.  The issue of reference states and
their appropriate choice is discussed at length in \cite{kassner01}.

The elastic equations must be supplemented with boundary conditions,
of which only the boundary conditions at the interface between the two
phases need to be specified here:
\begin{equation}
  \label{eq:boundfront}
  \sigma_{nn} = - p, \quad \sigma_{nt_1}= \sigma_{nt_2} = 0 \>,
\end{equation}
that is, the normal component of the stress is equal to the negative
external pressure, while its shear components along two independent
tangents $\tanga$ and $\tangb$ vanish.  By assuming a continuous
component $\sigma_{nn}$, the pressure jump at a curved interface due
to capillary overpressure is neglected, an approximation that is
well-justified under a wide range of conditions.  For a discussion of
the validity of this approximation as well as the others mentioned,
see \cite{durand98}.

Once the chemical potential has been computed, the stability of the
interface can be assessed, but in order to determine its evolution,
some dynamical law governing its motion must be stated.  If a particle
reservoir is present and the interface is rough, it is natural to
assume linear nonequilibrium kinetics. The driving force then is the
chemical potential difference itself, and the normal velocity $v_n$ of
the interface will be proportional to it: $v_n = - k_v \delta\mu$,
where $k_v$ is a mobility and the normal points from the solid into
the second phase.  For material transport by surface diffusion, the
driving force is the gradient of the chemical potential along the
surface instead, producing a surface current $j \propto -\nabla_s
\,\delta \mu$ ($\nabla_s$ is the surface gradient), which leads to a
dynamical law of the form
\begin{equation}
  \label{eq:vn_surfdiff}
  v_n =  D_s \,\Delta_s \delta \mu \>,
\end{equation}
where $\Delta_s$ is the Laplace-Beltrami operator on the surface and
$D_s$ a diffusion coefficient, assumed constant here.

Equations \eqref{eq:delta_mu} through \eqref{eq:vn_surfdiff}
constitute the continuum model for the ATG instability with transport
by surface diffusion, to which a phase-field model should converge in
the limit of asymptotically small interface width.

\section{Scalar-mobility phase-field model\label{sec:scalar}}

Before discussing the structure of previous phase-field models
attempting to capture surface diffusion dynamics, let us briefly
recall the phase-field model for nonconserved order parameter $\phi$.
This can be written \cite{kassner01}
\begin{equation}
  \label{eq:phidyn_nc}
 \frac{\partial \phi}{\partial t} =
\frac{k_v \gamma}{\rho_s} \bigg\{ \nabla ^2 \phi
-  \frac{1}{\epsilon^2} \Big[2 f'(\phi)
 + \frac{\epsilon}{3\gamma} h'(\phi) \potelast\Big]\bigg\}
\end{equation}
where $f(\phi) = \phi^2 (1-\phi)^2$ is the usual double-well potential
describing two-phase equilibrium and $h(\phi) = \phi^2 (3-2\phi)$ may
be interpreted as the local solid fraction, varying between 1
 in the solid phase (i.e., for $\phi=1$) and 0  in the
non-solid phase ($\phi=0$).  Primes denote derivatives with respect to the
argument.  $h(\phi)$ has the convenient property that it has extrema
at $\phi=0$ and $\phi=1$, the location of the minima of the
double-well potential.  $\potelast$ is given by
\begin{equation}
  \label{eq:defpotelast}
\potelast = \mus \sum_{i,j} \ud_{ij}^2 +
\frac{\lambda-\lamtil}{2} \Big(\sum_{k}\ud_{kk}\Big)^2 \>,
\end{equation}
where $\mus = E/[2(1+\nu)]$ is the shear modulus or first Lamé
constant, $\lambda = E\nu/[(1+\nu)(1-2\nu)]$ the second Lamé constant,
and $\lamtil$ is the bulk modulus of the non-solid phase (if this is
vacuum, $\lamtil=0$).

In the diffuse-interface formulation, the elastic equations take the
following form
\begin{eqnarray}
  \label{eq:elasteqs_ph}
  \sum_j \frac{\partial\ssigma_{ij}}{\partial x_j} &=& 0 \>, \nonumber\\
  \ssigma_{ij} &=& h(\phi) \,\sigma_{ij} - [1-h(\phi)]\, p\, \delta_{ij}  \>,
\end{eqnarray}
where $\sigma_{ij}$ is given by \eqref{eq:hooke}
and $p = p_0 - \lamtil \sum_k \ud_{kk}$.

The standard approach to a phase-field description of surface
diffusion, as proposed in \cite{cahn96,mahadevan99a,bhate00}, is then
to prepend the right hand side of Eq.~\eqref{eq:phidyn_nc} with a
differential operator corresponding to the divergence of a gradient
multiplied by a phase-field dependent mobility, i.e.,
Eq.~\eqref{eq:phidyn_nc} becomes replaced with
\begin{eqnarray}
  \label{eq:phidyn_c}
  \frac{\partial \phi}{\partial t} &=&
\nabla \cdot M \nabla  \frac{1}{\epsilon^2}\,
\delta\mu(\nabla^2\phi,\phi) \>,\nonumber\\
\delta\mu(\nabla^2\phi,\phi) &\equiv&
%\bigg\{ 
-\epsilon^2\nabla^2 \phi
+ 2 f'(\phi)
 + \frac{\epsilon}{3\gamma} h'(\phi) \potelast%\bigg\}
\>,\nonumber\\
\end{eqnarray}
where $M$ is a scalar function of either $\phi$ \cite{cahn96,bhate00} or
$\epsilon\nabla \phi$ \cite{mahadevan99a}, chosen such that the
mobility tends to zero far from the interface:
$M(\phi,\epsilon\nabla\phi)\to 0$ for $\phi\to 0$ and $\phi\to 1$.
The elastic equations remain unchanged, i.e., while the
nonconservative model is given by Eqs.~\eqref{eq:phidyn_nc} through
\eqref{eq:elasteqs_ph}, the conservative model is given by
Eqs.~\eqref{eq:defpotelast} through \eqref{eq:phidyn_c}.

At this point, a few remarks are in order.  First, the field $\phi$ is
the density of a conserved quantity by construction, since the right
hand side of \eqref{eq:phidyn_c} is written as a divergence.  This is
true for any (nonsingular) form of the mobility.  Second, $\delta\mu$
becomes zero for $\phi\to 0$ and $\phi\to 1$, meaning that there is no
diffusion in the bulk anyway.  One might therefore wonder whether it
is really necessary to choose a mobility that goes to zero in the
bulk.  The conservation law plus the absence of diffusion far from the
interface should suffice to restrict transport to diffusion along the
interface.  In fact, we shall see that essentially the same asymptotic
results are obtained no matter what the form of $M$, the only
conditions to be imposed being positivity (for almost all values of
$\phi$ or $\nabla\phi$) and boundedness.  It is just easier to derive
them if it is in addition assumed that $M$ vanishes in the bulk.  On
the other hand, it will turn out that if a restriction imposed by the
asymptotics is removed (or not yet satisfied in the temporal evolution
of the system), $M$ has to decay sufficiently fast inside the bulk for
the limit to make sense.  This may be relevant for the behaviour of
the model before it reaches its asymptotic state.

Finally, the issue at present is not so much whether the dynamics is
conservative but whether it does reduce to the sharp-interface model
of Sec.~\ref{sec:sharp_i_mod} in the limit of an asymptotically
vanishing interface thickness.  To investigate this, we have to
explicitly carry out the asymptotic analysis.

\subsection{Local coordinate system \label{sec:loccoord}}

The basic idea of the analysis is to expand all dynamical quantitities
in terms of the small parameter $\epsilon$ describing the interface
thickness, to solve for the phase field and to use the solution to
eliminate its explicit appearance from the equations.  To this end,
the domain of definition of the field is divided into an outer region,
where gradients of the fields can be considered to be of order one and
an inner region (close to the interface), where these gradients are
of order $1/\epsilon$.  The expansion in powers of $\epsilon$ is
rather straightforward in the outer domain,
Eqs.~\eqref{eq:defpotelast} through \eqref{eq:phidyn_c} can be taken
directly as a starting point.  As to the inner domain (and its
matching with the outer region), it is useful to first transform to
coordinates adapted to its geometry.  Therefore, a coordinate $r$ is
introduced that is orthogonal to the interface (defined as the level
set corresponding to $\phi(x,y,z,t)=1/2$) and two coordinates $\s$ and
$\uc$ that are tangential to it.  $r$ is the signed distance from the
interface and will be rescaled by a stretching transformation
$r=\epsilon\rho$ to make explicit the $\epsilon$ dependence for the
expansion.  Inner and outer solutions must satisfy certain matching
conditions due to the requirement that they agree in the combined
limit $\epsilon\to 0$, $\rho\to\pm\infty$, $r\to 0$.  These conditions
are given in App.~\ref{sec:matching}.

To obtain a basis for our local coordinate system, we first write
\begin{equation}
  \label{eq:posvec_loc}
 \rvec = \Rvec(\s,\uc)+r\, \norm(\s,\uc)
\end{equation}
where $\rvec$ is the position vector of a point near the interface,
$\Rvec$ the position of the interface itself, and $\norm$ the normal
vector on it (oriented the same way as in the sharp-interface model,
i.e., pointing out of the solid).  Next, we require the coordinates
$\s$ and $\uc$ parametrizing the interface to be both orthogonal and
arclength-like, meaning that the tangential vectors $\tanga\equiv
\partial\Rvec/\partial \s$, $\tangb\equiv \partial\Rvec/\partial \uc$
are orthogonal and that they are unit vectors.  It is shown in the
App.~\ref{sec:loc_coord} that the requirement of both coordinates
being arclengths is less innocent than it seems -- it implies that the
coordinate lines are geodesics, which is a strong restriction.  (On a
planar surface, only cartesian coordinate systems satisfy this
condition.)  It may be worthwhile to point out a difference between
the two- and three-dimensional cases here.  In two dimensions, the
local coordinate system can essentially always be chosen global in the
tangential coordinate, for which we may take the arclength.  Locality
is only necessary in $r$ (this coordinate may turn singular for a
curved interface, when the distance from the latter reaches the radius
of curvature).  In three dimensions, an interface with a nontrivial
topology does not permit a description by a single coordinate patch,
so in general the coordinates $\s$ and $\uc$ have to be local, too.
However, for the asymptotic analysis, we never need to compare
far-away points along the interface, so a coordinate patch that is
valid in the neighbourhood of a point $\Rvec(\s_0,\uc_0)$ is quite
sufficient, and we can choose, without loss of generality, a geodesic
coordinate system, despite the fact that this will not necessarily be
global even on a curved interface with the topology of the plane.

Given the coordinates, it is a trivial matter to write down a
coordinate basis
\begin{align}
 \label{eq:covar_bas}
\Er  &\equiv \frac{\partial\rvec}{\partial r} = \norm(\s,\uc)  \>, 
  \nonumber\\
\Es  &\equiv \frac{\partial\rvec}{\partial \s} =
\frac{\partial\Rvec}{\partial \s}
+ r \frac{\partial\norm}{\partial \s} =  \left(1+r \kappa_1\right)\tanga
+ r\tau_1\tangb \>, \\
\Eu  &\equiv \frac{\partial\rvec}{\partial \uc} =
\frac{\partial\Rvec}{\partial \uc}
+ r \frac{\partial\norm}{\partial \uc} =  \left(1+r \kappa_2\right)\tangb
-r \tau_2\tanga \>.
\nonumber
\end{align}
The second equality in each of these equations, i.e., the
representation of the coordinate basis in terms of the (right-handed)
orthonormal basis ($\norm$,$\tanga$,$\tangb$) is derived in
App.~\ref{sec:loc_coord}. $\kappa_1$ and $\kappa_2$ are the normal
curvatures of the curves $\uc={\rm const.}$ and $\s={\rm const.}$,
respectively.  Because the geodesic curvatures of these are zero, they
are identical to the total spatial curvatures. Their sign is chosen
such that a protrusion has positive curvature, i.e., a sphere would
have positive curvatures, opposite to the traditional definition in
mathematics.  $\tau_1$ and $\tau_2$ are the torsions of these curves,
and in App.~\ref{sec:loc_coord} it is shown that
$\tau_1=-\tau_2\equiv \tau$.

Note that the basis vectors $\Es$ and $\Eu$ are guaranteed to be
orthogonal to each other only for $r=0$, i.e., {\em on} the interface.
It would certainly be possible to define coordinates that are
orthogonal in a whole neighbourhood of the interface point under
consideration, but the additional effort would not be justified by the
minor simplifications afforded.  What will become important in the
following, however, is that $\Er$ {\em is} orthogonal to the two other
basis vectors.  Moreover, $\Er$ has unit length for all values of $r$,
while the lengths of the two other basis vectors vary off the
interface (i.e., for $r\ne0$).

For many of the calculations to follow it is convenient to stick to
the nonnormalized coordinate basis most of the time and to use the
orthonormal basis only at the very end, when the dependence on simply
interpretable quantities such as the curvatures is to be exhibited.
This enables rather compact representations of most expressions. From
\eqref{eq:covar_bas}, we first obtain the metric coefficients
$g_{\alpha\beta}={\mathcal E}_\alpha {\mathcal E}_\beta$, where
$\alpha,\beta \in \left\{r,\s,\uc\right\}$. The metric tensor then
reads
\begin{align}
\label{eq:metric}
& \left(g_{\alpha\beta}\right) =  {\mathbf g} \nonumber \\
& =
 \left(\begin{array}{ccc}
1 &              0                 &          0                       \\
0 & (1+r \kappa_1)^2 + r^2 \tau^2  & 2 r \tau + r^2 (\kappa_1+\kappa_2)\tau\\
0 & 2 r \tau + r^2 (\kappa_1+\kappa_2)\tau  & (1+r \kappa_2)^2 + r^2 \tau^2
\end{array} \right) \>,
\end{align}
its determinant is
\begin{equation}
g \equiv \det{\mathbf g} =
\left[(1+ r \kappa_1) (1+ r \kappa_2) - r^2 \tau^2\right]^2 \>,
\end{equation}
and the contravariant components of the metric tensor are obtained as
\begin{align}
\label{eq:inv_metric}
& \left(g^{\alpha\beta}\right) =    {\mathbf g}^{-1}  \nonumber \\
& = \frac{1}{g} \left(\begin{array}{ccc}
g &              0                 &          0                       \\
0 & (1+r \kappa_2)^2 + r^2 \tau^2  & -2 r \tau - r^2 (\kappa_1+\kappa_2)\tau    \\
0 & -2 r \tau - r^2 (\kappa_1+\kappa_2)\tau  & (1+r \kappa_1)^2 + r^2 \tau^2
\end{array} \right) .\nonumber\\
\end{align}

From now on, we use the Einstein summation convention for pairs of
covariant and contravariant indices.  The vectors of the reciprocal
basis are obtained from $\Ehalph = g^{\alpha\beta} \Ebet$:
\begin{align}
 \label{eq:contravar_bas}
\Ehr  &= \nabla r = \norm(\s,\uc)  \>,    \nonumber\\
\Ehs  &= \nabla \s = \frac{1}{\sqrt{g}} \left(\left(1+r \kappa_2\right)\tanga
- r\tau\tangb\right) \>, \\
\Ehu  &= \nabla \uc = \frac{1}{\sqrt{g}} \left(\left(1+r \kappa_1\right)\tangb
-r \tau\tanga \right)\>.
\nonumber
\end{align}

The gradient and divergence read
\begin{align}
\nabla &= \Ehalph \partial_\alpha \>, \label{eq:nabla}\\
\nabla\cdot \Avec &= \frac{1}{\sqrt{g}} \partial_\alpha
\left(\sqrt{g} g^{\alpha\beta} A_\beta\right) \>.
\label{eq:divergence}
\end{align}

In the following, we will denote inner quantities by the uppercase
letter corresponding to the lowercase letter denoting the outer quantity,
whenever this is meaningful.

Since the interface will move in general and the coordinates $r$,
$\s$, $\uc$ are defined with respect to the interface,
there is also a transformation rule for the time derivative:
\begin{equation}
\partial_t f(x,y,z,t) = \partial_t F(r,\s,\uc,t) - 
\wel \nabla F(r,\s,\uc,t) \>,
\label{eq:timederiv}
\end{equation}
where $\wel(\s,\uc,t)$ is the
%$=\wel(x(\s,\uc),y(\s,\uc),z(\s,\uc),t)$ 
interface velocity. Equation \eqref{eq:timederiv} exhibits that the
time derivative in the comoving frame is a material derivative. In
order to formulate the matching conditions concisely, we will
occasionally also write the outer fields as functions of the variables
$r$, $\s$, and $\uc$ (without changing their naming letter, thus in this
case adhering to the physicists' convention of using a letter for a
physical quantity rather than a mathematical function).

\subsection{Inner equations}

To render the scales of the different terms more visible, we rewrite Eqs.~\eqref{eq:nabla}
and \eqref{eq:divergence} in the (still exact) form
\begin{align}
\nabla &= \frac{1}{\epsilon} \,\norm\, \partial_\rho
+ \Ehbalph \partial_\balpha \>, \label{eq:nablaeps}\\
\nabla\cdot \Avec &=
  \frac{1}{\sqrt{g}} \left(\frac{1}{\epsilon}\,\partial_\rho \sqrt{g} A_r +
 \partial_\balpha
\sqrt{g} g^{\balpha\bbeta} A_\bbeta \right) \>,
\label{eq:divergenceeps}
\end{align}
where sub- and superscripts carrying an overbar are taken from the set
$\{\s,\uc\}$ only in the summation.

%Denoting inner quantities by capital letters and 
Assuming, without loss of generality, that the tangential velocity of
the interface vanishes, Eq.~\eqref{eq:phidyn_c} takes the following
form
\begin{eqnarray}
  \label{eq:phidyn_c_inn}
  \partial_t \Phi -\frac{1}{\epsilon} v_n \partial_\rho \Phi &=&
\nabla \cdot M \nabla \frac{1}{\epsilon^2}\, 
\delta\mu(\nabla^2\Phi,\Phi)\>,\nonumber\\
 \delta\mu(\nabla^2\Phi,\Phi) &=&
%\bigg\{ 
-\frac{1}{\sqrt{g}}\,\partial_\rho \sqrt{g}\,\partial_\rho
 \Phi \nonumber\\
 && - \epsilon^2\frac{1}{\sqrt{g}}\,\partial_\balpha
 \sqrt{g} g^{\balpha\bbeta}
 \partial_\bbeta \Phi \nonumber\\
&& + 2 f'(\Phi)
 + \frac{\epsilon}{3\gamma} h'(\Phi) \potelast%\bigg\}
\>,
\end{eqnarray}
with %$v_n$ denoting the normal velocity and
\begin{align}
  \label{eq:diffop1_inn}
\nabla \cdot M \nabla &= \frac{1}{\epsilon^2} \frac{1}{\sqrt{g}}\partial_\rho
 \sqrt{g} M \partial_\rho   + \frac{1}{\sqrt{g}}\partial_\balpha \sqrt{g}
 g^{\balpha\bbeta} M \partial_\bbeta\>.
\end{align}
Hence, the leading term of the inner equation \eqref{eq:phidyn_c_inn}
with the differential operator given by \eqref{eq:diffop1_inn} is of
order $\epsilon^{-4}$.

The mechanical equilibrium condition \eqref{eq:elasteqs_ph} becomes
\begin{eqnarray}
  \label{eq:elasteq_inn}
 \nabla \Ssigma &=& \frac{1}{\sqrt{g}}\, \bigg(\frac{1}{\epsilon}\,\partial_\rho \sqrt{g}\,
 \Ssigma_{\gamma r} \Ehgam  \nonumber\\
&& \mbox{} + \partial_\balpha \sqrt{g}\, g^{\balpha\bbeta}
\Ssigma_{\gamma \bbeta} \Ehgam \bigg) = 0 \>.
\end{eqnarray}

\subsection{Expansions, matched asymptotic analysis \label{sec:expansions}}

To solve the outer and inner equations successively, we expand the
fields in both the outer and inner domains in powers of $\epsilon$
\begin{eqnarray}
\phi(x,y,z,t) &=& \phinul(x,y,z,t) + \epsilon\, \phione(x,y,z,t)
\nonumber\\
 && \mbox{} + \epsilon^2\phitwo(x,y,z,t)... \>,
           \label{phioutexp} \\
u_{ij}(x,y,z,t) &=& \uijnul(x,y,z,t) +
           \epsilon \uijone(x,y,z,t) \nonumber\\
 && \mbox{}  + \epsilon^2 \uijtwo(x,y,z,t)... \>,
           \label{uijoutexp}
\end{eqnarray}
and
\begin{eqnarray}
\Phi(r,\s,\uc,t) &=& \Phinul(r,\s,\uc,t) + \epsilon\,
\Phione(r,\s,\uc,t)
\nonumber\\
 && \mbox{} + \epsilon^2\Phitwo(r,\s,\uc,t)... \>,
           \label{phiinnexp} \\
U_{\alpha\beta}(r,\s,\uc,t) &=& \Uabnul(r,\s,\uc,t) +
           \epsilon \Uabone(r,\s,\uc,t) \nonumber\\
 && \mbox{}  + \epsilon^2 \Uabtwo(r,\s,\uc,t)... \>.
           \label{uabinnexp}
\end{eqnarray}
The transformation from cartesian tensor components (subscripts $i,j$)
to curvilinear ones (sub- and superscripts $\alpha,\beta$) is carried
out using standard rules of tensor calculus.  Details can be found in
App.~\ref{sec:sl_elastic}.  From the expansions of the strain fields,
similar expansions (not written out here) follow for the stress
fields.  Moreover, the basic field variables for which the elastic
equations, the Lamé equations, ultimately constitute a closed
set, are the displacement vectors $u_i$ or $U_\alpha$.  A peculiarity
of phase-field formulations of elasticity is then that the fields
$U_\alpha^{(0)}$ are independent of $\rho$ \cite{kassner01}, i.e.,
continuous across the interface, because components of the strain
tensor contain derivatives $\partial_\rho U_\alpha/\epsilon$, but
their leading order is $\epsilon^{\,0}$, not $\epsilon^{-1}$.  Again,
details are presented in App.~\ref{sec:sl_elastic}.

The asymptotic analysis of the model
%Eqs.~\eqref{eq:elasteqs_ph} through \eqref{eq:phidyn_c}
may be divided into two natural subproblems: analysis of the equation
for the phase field and analysis of the mechanical equations; the
latter depend on the former only via the solution for the phase
field. %$\phi$.
Since the focus of this article is on how to capture surface diffusion
dynamics correctly, only the first subproblem will be treated in the
main text, a strategy that will hopefully facilitate the application
of the results to other models, such as phase-field simulations of
electromigration, for which they should possess relevance, too.  The
derivation of the sharp-interface limit of the mechanical equations is
a straightforward extension of the two-dimensional case, albeit with
one or two additional subtle points.  Because the three-dimensional
case does not seem to have been discussed in the literature before,
this is done in App.~\ref{sec:sl_elastic}. Another motivation to carry
this part of the analysis out explicitly is to present, as a case in
point, an operational phase-field model for a complete physical
system.

\subsubsection{Leading order}

The leading-order outer equation for $\phi$ is
\begin{equation}
\label{eq:scalout_0}
 \nabla \cdot M \nabla f'(\phinul) = 0\>,
\end{equation}
which is to be supplemented with the boundary conditions $\phinul=1$
and $\phinul=0$ at infinity in the regions where the system is solid
and non-solid, respectively.  If we regard \eqref{eq:scalout_0} as a
partial differential equation for the function $f'(\phinul)$ (rather
than for $\phinul$ itself), this boundary condition translates into
$f'(\phinul)\to 0$ as $\abs{\rvec}\to \infty$, which may be seen
immediately from the explicit form of $f'(\phi)$, given in
App.~\ref{sec:collection}.  The new boundary condition is valid
everywhere at infinity except possibly in a region with size of order
$\epsilon$.  For general $M(\phi,\epsilon\nabla\phi)$, the partial
differential equation (\ref{eq:scalout_0}) is of course nonlinear.
Nevertheless, it can be shown to have the unique solution $f'(\phinul)
= 0$, if $M$ is positive everywhere, except possibly on a set of
measure zero.

To see this, multiply Eq.~\eqref{eq:scalout_0} by $f'(\phinul)$,
integrate over all of space and use Gauss's theorem:
\begin{align}
\label{eq:uniqueness}
 0 &= \int d^3 x f'(\phinul) \nabla \cdot M \nabla f'(\phinul)
 \nonumber\\
 &= -\int d^3 x  M \left[\nabla f'(\phinul)\right]^2\> + O(\epsilon)\>,
\end{align}
where the $O(\epsilon)$ stands for the surface integral at infinity.
If $M$ is positive almost everywhere, we immediately get
$f'(\phinul)=\hbox{const.}$, and the boundary conditions require the
constant to be zero.  This conclusion remains of course unchanged, if
$M$ becomes zero only when $\phinul$ is zero or one -- a standard
choice \cite{bhate00} is $M(\phi) \propto \phi^2 (1-\phi)^2$.

Hence, the unique solution to the leading-order outer problem is, if
we now consider it an equation for $\phinul$ again, $\phinul=1$ in
$\Omega^{-}$ and $\phinul=0$ in $\Omega^{+}$, where $\Omega^{\mp}$ are
those regions of space, separated by the interface(s), in which
$\lim_{\abs{\rvec}\to\infty} \phinul = 1$ and $0$, respectively.  The
solution $\phinul=\frac12$, still possible for the equation
interpreted as an equation for $f'(\phinul)$, is excluded by the
boundary conditions for $\phinul$.  (This argument presupposes that we
have no domains that are not connected with infinity.  For the
interior of a closed interface, the solution $\phinul=1/2$ would have
to be excluded by a stability argument or by making reference to
initial conditions.)

It is then seen by inspection that the outer equation is indeed solved
to {\em all\/} orders by the solution under discussion (because
$h'(0)=h'(1)=0$).  Therefore, we have $\phione\equiv0$,
$\phitwo\equiv0$, which provides us with partial boundary conditions
for the inner solutions $\Phione$, $\Phitwo$, and so on (see
App.~\ref{sec:matching}).  Moreover, only the inner problem needs to
be considered beyond the leading order.

Because $g=1+O(\epsilon)$, the leading-order inner problem becomes [see
Eqs.~\eqref{eq:phidyn_c_inn} and \eqref{eq:diffop1_inn}]
\begin{equation}
\label{eq:phydin_c_innlead}
\partial_\rho M(\Phinul) \partial_\rho \left[\partial_{\rho\rho} \Phinul -2
f'(\Phinul)\right]=0 \>,
\end{equation}
which can be integrated once to yield
\begin{equation}
\label{eq:phydin_c_innlead_int}
\partial_\rho \left[\partial_{\rho\rho} \Phinul -2
f'(\Phinul)\right] = \frac{c_1(\s,\uc)}{M(\Phinul)} \>,
\end{equation}
where $c_1(\s,\uc)$ is a function of integration.  It is here that we
have to follow different lines of arguments, depending on whether $M$
approaches zero for $\rho\to\pm\infty$, which is the case for the
mobilities assumed in \cite{cahn96,bhate00}, or whether it is just a
bounded (and possibly constant) function of $\phi$.  In the first
case, we may immediately conclude $c_1=0$, because the right hand side
of (\ref{eq:phydin_c_innlead_int}) must not diverge.  In the second
case, we obtain the same result by integrating
\eqref{eq:phydin_c_innlead_int} first and invoking the boundary
conditions:
\begin{equation}
\label{eq:phydin_c_innlead_int2}
\partial_{\rho\rho} \Phinul -2
f'(\Phinul) = c_1(\s,\uc)\int_0^\rho \frac{1}{M} d\rho + c_2(\s,\uc)\>.
\end{equation}
Since $M$ is bounded from above and positive, the integral will be
larger in magnitude than $\int_0^\rho 1/(\sup_\rho M) d\rho
=\rho/\sup_\rho M$, so the two terms on the right hand side of
Eq.~\eqref{eq:phydin_c_innlead_int2} are linearly independent.
The left hand side approaches zero for $\rho\to\pm\infty$ [the
argument will be made more rigorous below in the discussion of
$\Phione$], so both $c_1$ and $c_2$ must be equal to zero.  To argue
that $c_2$ is zero in the case where $M\to0$ for $\rho\to\pm\infty$, we
can proceed the same way, except that we have already gotten rid of
the term containing $c_1$, so the right hand side of
\eqref{eq:phydin_c_innlead_int2} is $c_2$ only.

To summarize, the leading-order inner equation results in
\begin{equation}
\label{eq:phydin_c_innlead_res}
\partial_{\rho\rho} \Phinul -2 f'(\Phinul) = 0\>,
\end{equation}
and this provides us with the solution $\Phinul= \nicefrac12
\,(1-\tanh\rho)$ as is shown in App.~\ref{sec:collection}.

\subsubsection{Next-to-leading order}

The next-to-leading order in Eq.~\eqref{eq:phidyn_c_inn} is the order
$\epsilon^{-3}$.  Since the differential operator in front of the
chemical potential is of order $\epsilon^{-2}$ and the chemical
potential multiplied by another factor $\epsilon^{-2}$, we must expand
$\delta\mu$ up to order $\epsilon$.  Equation
\eqref{eq:phydin_c_innlead_res} already tells us that
$\delta\mu^{(0)}=0$, so we obtain
\begin{equation}
\label{eq:phydin_c_innnext}
\partial_\rho M(\Phinul) \partial_\rho \delta\mu^{(1)}=0 \>,
\end{equation}
from which we get
\begin{equation}
\label{eq:phydin_c_innnext1}
 \partial_\rho \delta\mu^{(1)}= \frac{d_1(\s,\uc)}{M(\Phinul)} \>.
\end{equation}
As before, we can immediately conclude from this that $d_1=0$, if we
assume $M(\Phinul)\to 0$ for $\Phinul\to 0,1$. For arbitrary but bounded
$M$, we invoke the matching conditions (see App.~\ref{sec:matching})
\begin{equation}
\lim_{\rho\to\pm\infty} \partial_\rho \delta\mu^{(1)} = \partial_r
 \delta\mu^{(0)}_{\rm out}\vert_{\pm 0} = 0
\end{equation}
to obtain the same result (where for once we have denoted an outer
quantity by a subscript ''out'')

Integrating once more with respect to $\rho$ and writing out
$\delta\mu^{(1)}$, we have
\begin{eqnarray}
\label{eq:deltamu1res}
  \delta\mu^{(1)} &=& -\partial_{\rho\rho} \Phione - (\kappa_1+\kappa_2)
 \partial_\rho \Phinul \nonumber\\
  &&\mbox{}+ 2 f''(\Phinul)\,\Phione + \frac{1}{3\gamma} h'(\Phinul) 
\potelastnul
= d_2(\s,\uc) \>.\nonumber\\
\end{eqnarray} 
Up to this point, there is agreement between this and preceding
asymptotic analyses \cite{mahadevan99a}, if not in all details
of the procedure, so at least in the results (assuming the appropriate
replacement of the interaction with the electric field in the
electromigration system by that with elastic strain and simplifying
our result to the two-dimensional case).

Let us now try to determine the function of integration $d_2(\s,\uc)$.
A priori, there is no reason to use a procedure different from what we
have done before.  We know the limiting values for $\rho\to\pm\infty$
for three of the five terms on the right hand side~of \eqref{eq:deltamu1res}:
$\lim_{\rho\to\pm\infty} \partial_\rho \Phinul = 0$ [which follows
from either the matching conditions or by inspection of the solution
\eqref{eq:solution_Phi0}], $\lim_{\rho\to\pm\infty} h'(\Phinul) = 0$
[because $h'(0)=h'(1)=0$], $\lim_{\rho\to\pm\infty} d_2(\s,\uc) =
d_2(\s,\uc)$ (because $d_2$ is independent of $\rho$).  Moreover, from
the matching conditions, we obtain the limit for $\Phione$
\begin{eqnarray}
\label{asympt_phione}
\Phione \sim \rho\phi'^{(0)}(\pm0)+\phione(\pm0) 
&=& \phione(\pm0) = 0 \nonumber\\
&&(\rho\to\pm\infty) \>.
\end{eqnarray}
The second equality follows from the fact that $\phinul=0$ or
$\phinul=1$, hence its derivative with respect to $r$ vanishes on both sides of
the interface; the third equality is a consequence of the fact that
$\phinul$ solves the outer equation to all orders and hence
$\phione\equiv 0$.

With four of the five terms in \eqref{eq:deltamu1res} having a
definite limit, we may conclude that the fifth must have a limit
as well and obtain
\begin{equation}
\label{eq:limitd2phi1}
\lim_{\rho\to\pm\infty} -\partial_{\rho\rho} \Phione = d_2(\s,\uc)\>.
\end{equation}
But if this limit exists, it cannot be different from zero:
transforming to $\xi=1/\rho$, we see that $\partial_{\rho\rho} \Phione
= \big(\xi^2 \partial_\xi \big)^2 \Phione$, which implies the
asymptotic behaviour $\Phione\sim -d_2/2\xi^2\>\> (\xi\to 0)$ and
hence the divergence of $\Phione$ as $ -d_2 \rho^2/2$, if $d_2\ne 0$.
(The same kind of argument can be used to show that the left hand side
of Eq.~\eqref{eq:phydin_c_innlead_int2} goes to zero, even though the
matching conditions do not provide a direct expression for
$\lim_{\rho\to\pm\infty}\partial_{\rho\rho} \Phinul$.)

The upshot of these detailed considerations is that 
\begin{equation}
  \label{eq:mu1eq0}
d_2(\s,\uc) = 0 \quad \Rightarrow \quad \delta\mu^{(1)} = 0. 
\end{equation}
Previous treatments of the problem did not enter into these
considerations.  Instead, one of the two following equivalent approaches
was chosen.  Either, Eq.~\eqref{eq:deltamu1res} was interpreted as a
linear inhomogeneous differential equation for $\Phione$ and
Fredholm's alternative invoked.  Since the appearing linear
operator
\begin{equation}
  \label{eq:linop}
 \linopl = \partial_{\rho\rho}-2 f''(\Phinul)
\end{equation} 
is hermitean, we know (from taking the derivative of
Eq.~(\ref{eq:phydin_c_innlead_res}) w.r.t.~$\rho$) that
$\partial_\rho\Phinul$ is a solution to the adjoint homogeneous
equation. The inhomogeneity of the differential equation must be
orthogonal to this solution.  Or else, Fredholm's alternative was not
mentioned, the equation was simply multiplied by
$\partial_\rho\Phinul$, integrated, and it was shown via integration
by parts that the terms containing $\Phione$ disappear. Of course,
this is the same thing.

To exploit the approach optimally, we use the relationship
$\partial_\rho\Phinul=- h'(\Phinul)/3$ shown in
App.~\ref{sec:collection}.  This way we obtain from
\eqref{eq:deltamu1res}
\begin{align}
\label{eq:fredholm_1}
& -\int_{-\infty}^{\infty} \left[\kappa_1+\kappa_2 
+\frac{1}{\gamma}\potelastnul\right] \left(\partial_\rho 
\Phinul\right)^2 d\rho \nonumber\\
&= \int_{-\infty}^{\infty} \partial_\rho \Phinul d_2(\s,\uc) d\rho = -
d_2(\s,\uc)
\end{align}
or
\begin{equation}
  \label{eq:fredholm_2}
   d_2(\s,\uc) =  \frac13 \left(\kappa_1+\kappa_2 
+ \potelastnulbar/\gamma\right) \>,
\end{equation}
where we have used \eqref{eq:intdphirho2} and introduced $\potelastnulbar
 =\int_{-\infty}^{\infty}\potelastnul \Big(\partial_\rho \Phinul\Big)^2 d\rho/
\int_{-\infty}^{\infty}  \Big(\partial_\rho \Phinul\Big)^2 d\rho$.

Both Eqs.~\eqref{eq:mu1eq0} and \eqref{eq:fredholm_2} were derived by
valid methods, therefore they should both hold true.  Nevertheless, as
we shall see shortly, Eq.~\eqref{eq:mu1eq0} is a quite undesirable
result.  This may be the deeper reason why it was so far overlooked
and only the analog of Eq.~\eqref{eq:fredholm_2} derived.  When
Eq.~\eqref{eq:mu1eq0} is inserted in \eqref{eq:fredholm_2} it imposes a
relationship between the elastic state of the material (represented by
$\potelastnulbar$) and the curvature. (In models, where the
interaction term is quadratic in $\epsilon$ \cite{mahadevan99a}, it
would even impose the restriction $\kappa_1+\kappa_2=0$.)

\subsubsection{Higher orders}
To see that the model would indeed work if we did not have the
restriction \eqref{eq:mu1eq0}, let us consider the equations at the
next two orders, ignoring for the time being the result $d_2=0$.
Since both $\delta\mu^{(0)} (=0)$ and $\delta\mu^{(1)}$ are
independent of $\rho$, the first term of the operator
\eqref{eq:diffop1_inn} does not produce any contribution from these
terms in (\ref{eq:phidyn_c_inn}), and the order $\epsilon^{-2}$
equation reads
\begin{equation}
  \label{eq:phydin_c_inn2}
  \partial_\rho M \partial_\rho \delta\mu^{(2)}
   +\partial_\balpha  g^{\balpha\bbeta} M \partial_\bbeta\,\delta\mu^{(0)} 
= 0 \>,
\end{equation}
where we can immediately drop the second term, because of
$\delta\mu^{(0)} =0$.  After two integrations this becomes
\begin{equation}
  \label{eq:inn2_int}
 \delta\mu^{(2)} =
  e_1(\s,\uc)\int_0^\rho \frac{1}{M} d\rho + e_2(\s,\uc)\> \>.
\end{equation}
If $M\to 0$ for $\rho\to\pm\infty$, we immediately find $ e_1(\s,\uc)=0$.  In the general case, we use the matching conditions [see \eqref{eq:asymrels3}]
\begin{eqnarray}
  \label{eq:matchmu2}
  \delta\mu^{(2)} &\sim& \frac12 \rho^2 \,\partial_{rr} 
\delta\mu^{(0)}_{\rm out}\vert_{r=\pm0}
+ \rho \,\partial_{r} \delta\mu^{(1)}_{\rm out}\vert_{r=\pm0} \nonumber\\
&&\mbox{} +  \delta\mu^{(2)}_{\rm out}\vert_{r=\pm0} \>.
\end{eqnarray}
From Eq.~\eqref{eq:phidyn_c}, we gather that an expansion of
$\delta\mu_{\rm out}$ in powers of $\epsilon$ will contain three types
of terms, the first of which have the form $\nabla^2 \phi^{(k)}$
($k=0,1,\dots$), while the second contain factors $\phi^{(k)}$
($k=1,2, \dots$), coming from an expansion of $f'(\phi)$ or $h'(\phi)$
about $\phinul$, and the third include either $f'(\phinul)$ or
$h'(\phinul) \potelast^{(k)}$ ($k=0,1,\dots$).  All of these terms
vanish, because $\phi^{(k)}=0$ for $k>0$ and because $f'(\phinul) =
h'(\phinul) = 0$.  This is simply a consequence of the fact that the
outer equation is solved exactly by $\phinul =0$ and $\phinul = 1$.
The ``chemical potential'' appearing in the phase-field equations
needs to be related to the true, i.e., sharp-interface chemical
potential only at the interface.  In the outer domain, it is zero.  We
can then conclude from \eqref{eq:matchmu2} that $e_1(\s,\uc)=0$ (of
course $e_2(\s,\uc)=0$, too, but we shall not make use of this
result).

Given that $\delta\mu^{(2)}$ is independent of $\rho$, the inner
equation at order $\epsilon^{-1}$ takes the form
\begin{equation}
  \label{eq:phydin_c_inn3}
  -v_n \partial_\rho \Phinul =
\partial_\rho M \partial_\rho \delta\mu^{(3)}
   +\partial_\balpha  g^{\balpha\bbeta} M \partial_\bbeta\,\delta\mu^{(1)} \>.
\end{equation}
Because the surface metric $g^{\balpha\bbeta}=\delta_{\balpha\bbeta}$
to lowest order in $\epsilon$, we find after integration over $\rho$
($v_n$ does not depend on $\rho$)
\begin{eqnarray}
   \label{eq:veloc_scalmodprel}
   v_n &=& \left(\partial_{\s\s}+\partial_{\uc\uc}\right)\delta\mu^{(1)}
\int_{-\infty}^{\infty} M(\Phinul) d\rho \nonumber\\
&&\mbox{}+  M(\Phinul) \partial_\rho\delta\mu^{(3)}\vert_{-\infty}^{\infty}\>. 
\end{eqnarray}
Here we can drop the second term on the right hand side, if
$\lim_{\rho\to\pm\infty}M(\Phinul)=0$.  Formally setting
$\int_{-\infty}^{\infty} M(\Phinul) d\rho = 3 M^\ast \gamma$ and using
\eqref{eq:fredholm_2}, we arrive at
\begin{equation}
 \label{eq:veloc_scalmod}
 v_n =  M^\ast \left(\partial_{\s\s}+\partial_{\uc\uc}\right) 
\left[\gamma(\kappa_1+\kappa_2) + \potelastnulbar\right] \>.
\end{equation}
It is shown in App.~\ref{sec:sl_elastic} that $\potelastnulbar$ is the
correct sharp-interface limit of the elastic part of the chemical
potential in Eq.~\eqref{eq:delta_mu}, hence \eqref{eq:veloc_scalmod}
reproduces the sharp-interface limit \eqref{eq:vn_surfdiff}, as the
surface Laplacian in geodesic orthogonal coordinates is given by
\begin{equation}
 \label{eq:laplace_beltrami} 
\Delta_s = \partial_{\s\s}+\partial_{\uc\uc} \>.
\end{equation}

Finally, $M^\ast$ would be infinite for positive functions
$M(\Phinul)$ that do not reduce to zero for $\rho\to\pm\infty$;
therefore, in the end we would indeed have to require $M(\Phinul)$ to
decay far from the interface, if $\delta\mu^{(1)}$ were different from
zero.  In reality, we do not just have \eqref{eq:veloc_scalmod}, the
equation we want, but in addition Eq.~\eqref{eq:mu1eq0}, requiring
$\delta\mu^{(1)}=0$ and, consequently
\begin{equation}
 \label{eq:veloceq0}
 v_n =  0 \>.
\end{equation}
Thus we are led to the conclusion that the asymptotics of the
phase-field model produces the correct sharp-interface limit {\em only
  at equilibrium}, i.e., when the normal velocity vanishes.  Of
course, one may ask what will happen, if we prepare the system in an
initial state, where Eq.~\eqref{eq:phidyn_c} requires the phase field
to change. Then $v_n$ will necessarily be different from zero.  The
answer is that in such a case the phase field will not follow its
asymptotic dynamics yet.  A similar phenomenon happens when a
phase-field simulation is started with an initial interface perturbed
by white noise.  Since the asymptotics of the phase-field equations
require curvatures to be smaller than $1/\epsilon$, the initial stage
of the dynamics where larger curvatures are present, will not be
governed by these asymptotics.  But the asymptotic behaviour is
sufficiently robust to keep that initial stage short.  It is tempting
to speculate that when conditions are such that \eqref{eq:veloceq0}
does not hold yet, the phase-field model discussed here will satisfy all
less restrictive conditions such as \eqref{eq:veloc_scalmod} already.
%while only developing towards the realization of \eqref{eq:veloceq0}.
Then the model would be applicable during the period where the
influence of condition \eqref{eq:mu1eq0} leading to
\eqref{eq:veloceq0} is still small.  However, it should be clear that
without a theoretical estimate of the error in this
not fully asymptotic state, the model can hardly be considered
quantitative.  Condition \eqref{eq:mu1eq0} should be expected to have
a stabilizing influence on the interface, rendering growth of
instabilities less rapid than in the sharp-interface model.
%exhibit a dynamical behaviour characterized by 

At first sight, the deep quench case considered in \cite{cahn96} seems
to escape an analogical consequence, because there the matching
procedure is performed on a finite $\rho$ domain and hence an argument
leading to the restriction (\ref{eq:mu1eq0}) appears to be missing.
Indeed, instead of $f(\phi)$ the model has a quadratic potential
$\tilde{f}(\phi) = (1-\phi^2)/2$, leading to sinusoidal behaviour of
$\Phinul$, which takes its limiting values -1 and 1 for finite $\rho$
($=\pm \pi/2$) already.  However, closer inspection shows that the
problem has not disappeared, it just takes a hidden form.  The phase
field must be specified beyond the domain of thickness
$r=\orderof{\epsilon}$ where $\tilde{f}(\phi)\ne 0$, which means that
the definitions of $\tilde{f}'(\Phinul)$ $(=-\Phinul)$ and
$\tilde{f}''(\Phinul)$ $(=-1)$ have to be extended to include the
values $\Phinul = \pm 1$ into their domain.  To keep Eq.~(3.10b) of
\cite{cahn96} valid for arbitrarily large $\rho$, it is necessary to
set $\tilde{f}'(\pm 1) =0$, i.e., to impose jump discontinuities of
size 1 at the ends of the $\Phinul$ interval [-1,1].  This is evident,
as the stationary values of the phase field should correspond to
minima of the potential.  However, Eq.~(3.13b) of \cite{cahn96} then
becomes
\begin{equation}
 \mu(s,t) = -\Phione_{\rho\rho} - \kappa \Phinul_\rho + \tilde{f}''(\Phinul)
\,\Phione \>,
\label{eq:to_ignore}
\end{equation}
where the behaviour of the second derivative of the potential is
essentially given by $\tilde{f}''(\Phinul) = -1 + \delta(\Phinul-1) +
\delta(\Phinul+1)$.  In any case, with $\tilde{f}'(\Phinul)$
exhibiting only jump discontinuies, $\tilde{f}''(\Phinul)$ cannot have
stronger singularities than $\delta$ functions. The last term in
Eq.~(\ref{eq:to_ignore}) then becomes zero for $\abs{\rho}>\pi/2$,
because $\Phione$ approaches zero continuously for $\rho\to\pm \pi/2$
(and is identically 0 for $\abs{\rho}>\pi/2$).  Then the total right
hand side of the equation will approach zero for $\rho\to\pm\infty$,
again forcing the chemical potential $\mu(s,t)$ to be equal to zero.

It is instructive to note why the nonconservative model obtained when
\eqref{eq:phidyn_c} is replaced with \eqref{eq:phidyn_nc} does {\em not\/}
suffer from a similar difficulty.  In that model, the velocity is already
determined at the next-to leading order.  Instead of
\eqref{eq:deltamu1res}, we get 
\begin{eqnarray}
\label{eq:veloc_nc}
-v_n \partial_\rho \Phinul&=& 
\frac{k_v \gamma}{\rho_s} \bigg\{\partial_{\rho\rho} \Phione
 + (\kappa_1+\kappa_2) \partial_\rho \Phinul \nonumber\\
  &&\mbox{}- 2 f''(\Phinul)\,\Phione - \frac{1}{3\gamma} h'(\Phinul) 
\potelastnul\bigg\}
 \>. \nonumber\\
\end{eqnarray}
Again we may conclude that all the terms on the right hand side go to
zero as $\rho$ is sent to $\pm\infty$.  However, this does not lead to
any constraints, since the left hand side is $\rho$ dependent now and
goes to zero as well, satisfying the limit automatically, whereas in
the surface-diffusion case, it was a function of $\s$ and $\uc$ only
($d_2$) that could be concluded to be equal to zero.  So consideration
of the limit does not produce anything new here, and the only
procedure available to extract information on $v_n$ is to use
Fredholm's alternative which gives the correct sharp-interface limit.

In the case of the nonconservative model, the introduced chemical
potential functional is zero in the bulk just as in the conservative
case, but there are no restrictions on its variation near the
interface, where it acquires a form tending to a $\delta$ function in
the sharp-interface limit.  In the conservative model, this is
excluded by restrictions on the derivative of the chemical potential
with respect to $\rho$, meaning that the latter must be smooth across
the interface.  Since it is zero off the interface, it is zero on it
as well.  Due to this reason, the phase-field model strictly speaking
applies only to the equilibrium limit.  Far-from equilibrium dynamics
is not likely to be captured faithfully.

\section{Tensorial mobility\label{sec:tensorial}}
%Rectification of the differential operator

That the phase-field model given by Eqs.~\eqref{eq:defpotelast} through
\eqref{eq:phidyn_c} does not quite yield the correct asymptotics may
be traced back to the fact that the differential operator $\nabla\cdot
M \nabla$, prepended to the chemical potential in \eqref{eq:phidyn_c},
does not reduce to the surface Laplacian $\Delta_s$ in the asymptotic
limit.  In fact, the second term on the right hand side of
Eq.~\eqref{eq:diffop1_inn} {\em is} the Laplace-Beltrami operator on
the surface, but the first term, containing derivatives with respect
to $\rho$ is orders of magnitude larger, being preceded by a factor of
$1/\epsilon^2$.  As a consequence, the asymptotics must be
secured by the full solution of the equation rather than by both the
operator and the chemical potential converging to the desired
sharp-interface limits.

Realizing this property of the model, it seems natural to modify the
differential operator via introduction of an essentially tensorial
mobility.  Let us denote by
\begin{equation}
  \label{eq:normhat}
  \normhat = -\frac{\nabla\phi}{\abs{\nabla\phi}}
\end{equation}
the normal on the surface $\phi={\rm const.}$ (for $\phi=1/2$, we
have $\normhat=\norm$), then we expect the operator 
\begin{equation}
  \label{eq:diffop_proj}
  \nabla\cdot \left(1-\normhat:\normhat\right) \nabla
\end{equation}
to reduce to the surface Laplacian asymptotically. A colon is used to
designate a dyadic product, so $1-\normhat:\normhat$ is a projection
operator projecting onto the tangential plane of a level set of
$\phi$. Introducing the shorthand $\phikalph =
\partial_\alpha\phi$, we have $\nabla\phi = \Ehalph \phikalph$ and
\begin{align}
  \label{eq:eval_diffproj}
  \nabla\cdot &\left(1-\normhat:\normhat\right) \nabla = \nonumber\\
& \quad\quad \nabla\cdot \bigg(\Ehmu \partial_\mu - \frac{1}{(\nabla\phi)^2} 
\Ehalph \phikalph g^{\beta\mu} \phikbet \partial_\mu\bigg)  = \nonumber\\
& \quad\quad \frac{1}{\sqrt{g}} \partial_\nu \sqrt{g} g^{\nu\mu} 
\left(\partial_\mu-\frac{1}{(\nabla\phi)^2}\phikmu g^{\alpha\beta} \phikalph 
\partial_\beta\right)\>.
\end{align}
The third expression is obtained from the second applying the
divergence operator \eqref{eq:divergence} and renaming $\alpha\to\mu$,
$\beta\to\alpha$, $\mu\to\beta$ in the three pairs of ``mute''
indices.

To expand this operator in powers of $\epsilon$ in the inner domain,
we use the fact that $g^{\rho\balpha}=0$ and that 
\begin{eqnarray}
  \label{eq:nabphisq}
  (\nabla\Phi)^2 &\equiv& \frac{1}{\epsilon^2} (\tnabla\Phi)^2 \>, \nonumber\\
  (\tnabla\Phi)^2 &=&  \epsilon^2 \Phikalph g^{\alpha\beta} \Phikbet =
         \Phikrho^2 +  \epsilon^2 \Phikbalph g^{\balpha\bbeta} \Phikbbet \>. 
\end{eqnarray}
Inserting this into \eqref{eq:eval_diffproj} and carrying the
expansion to formal order $\epsilon^0$, we find first that the order
$\epsilon^{-2}$ terms (containing two derivatives with respect to
$\rho$) cancel each other.  The remainder reads
\begin{align}
 \label{eq:diffprolorder1}
 \nabla\cdot &\left(1-\normhat:\normhat\right) \nabla = \nonumber\\
&\frac{1}{\sqrt{g}}\,\partial_\rho \sqrt{g}\left(
\frac{\Phikbalph g^{\balpha\bbeta} \Phikbbet}{(\tnabla\Phi)^2} \partial_\rho
-\frac{\Phikrho g^{\balpha\bbeta} 
\Phikbalph}{(\tnabla\Phi)^2} \partial_\bbeta\right) \nonumber\\
 &+ \frac{1}{\sqrt{g}}\,\partial_\bnu
 \sqrt{g} g^{\bnu\bmu} \left(\partial_\bmu
-\frac{\Phikrho \Phikbmu}{(\tnabla\Phi)^2}
 \partial_\rho\right) + O(\epsilon)\>,
\end{align}
and this expression still contains derivatives with respect to $\rho$.
However, {\em if} the leading-order solution $\Phinul$ depends on
$\rho$ only, as it did in the last section, then all the derivatives
of $\Phi$ with respect to $\s$ and $\uc$ (i.e., with respect to the
variables marked by an overbar) are $O(\epsilon)$ at least, and
Eq.~\eqref{eq:diffprolorder1} reduces to $\nabla\cdot
(1-\normhat:\normhat) \nabla = \nicefrac{1}{\sqrt{g}}\,\partial_\bnu
\sqrt{g} g^{\bnu\bmu} \partial_\bmu + O(\epsilon)$, i.e., at leading
order the operator indeed becomes the Laplace-Beltrami operator on the
surface.

This then suggests to replace the phase-field equation   
\eqref{eq:phidyn_c} with 
\begin{equation}
  \label{eq:phidyn_c_tens} 
  \frac{\partial \phi}{\partial t} = M
\nabla \cdot \left(1-\normhat:\normhat\right) \nabla \,
\frac{1}{\epsilon^2}\delta\mu(\nabla^2\phi,\phi) \>,
\end{equation}
where $\delta\mu$ is unchanged from \eqref{eq:phidyn_c} but
$M$ is a constant mobility now. 

In this model, the equation for the velocity would appear at the
next-to leading order already and take the form
\begin{eqnarray}
\label{eq:app_veloc}
v_n \partial_\rho \Phinul&=& 
M \left(\partial_{\s\s}+\partial_{\uc\uc} \right) 
\bigg\{\partial_{\rho\rho} \Phione
 + (\kappa_1+\kappa_2) \partial_\rho \Phinul \nonumber\\
  &&\mbox{}- 2 f''(\Phinul)\,\Phione - \frac{1}{3\gamma} h'(\Phinul) 
\potelastnul\bigg\}
 \>. \nonumber\\
\end{eqnarray}
Because the operators $\linopl$ [defined in Eq.~(\ref{eq:linop})] and
$\partial_{\s\s}+\partial_{\uc\uc}$ commute, Fredholm's alternative
could be applied as in the nonconservative case.  This eliminates
$\Phione$ from the equation and produces the correct sharp-interface
limit.

In spite of this enjoyable state of affairs, model
\eqref{eq:phidyn_c_tens} fails much more miserably than
\eqref{eq:phidyn_c}. The reason is that the zeroth-order solution is
not unique.  In fact, the leading-order outer equation
\begin{equation}
\label{eq:outer0}
\nabla \cdot \left(1-\normhat^{(0)}:\normhat^{(0)}\right)
 \nabla f'(\phinul) = 0 
\end{equation} 
is solved by {\em any} (differentiable) function $\phinul$ 
satisfying the boundary conditions: obviously we have $\nabla
f'(\phinul) = f''(\phinul)\nabla\phinul$, whence
\begin{align}
  \label{eq:effect_proj}
  \normhat^{(0)}:\normhat^{(0)} \nabla f'(\phinul) &= -\normhat^{(0)}
f''(\phinul) \abs{\nabla\phinul} \nonumber\\
&= f''(\phinul) \nabla\phinul = \nabla f'(\phinul)\>,
\end{align}
which implies $(1-\normhat^{(0)}:\normhat^{(0)}) \nabla
f'(\phinul) = 0$ for all functions  $f'$ of $\phinul$.

It can be said that this model fails for reasons complementary to
those of the scalar model.  Whereas we had one equation too many in
that case, adding a constraint to the desired sharp-interface
dynamics, now we have one equation too few, as there is nothing in the
model fixing $\phinul$.  If we had the right $\phinul$, the tensorial
model would work perfectly. In particular, it would give the limiting
equations at lower order than the scalar model, which may have positive
implications for the numerical cost at a given desired level of
accuracy, as will be discussed in the conclusions.

\section{Model with correct asymptotics - modified tensorial
  mobility\label{sec:correct}}

In order to obtain a model not plagued by either of the disadvantages
of the two cases discussed, it appears that we have to combine ideas
from both.  While it is certainly desirable to have a differential
operator that itself approaches the surface Laplacian, it should do so
only for phase field functions that have the correct leading-order
profile.

This can be achieved by modifying $1-\normhat:\normhat$ into
\begin{equation}
\label{eq:defQ} 
Q\equiv 1-\epsilon^2 \frac{\nabla\phi:\nabla\phi}{4 f(\phi)} 
= 1 -  \frac{\epsilon^2 (\nabla\phi)^2}{4 f(\phi)} \normhat:\normhat\>.
\end{equation}  

If we replace the projection operator in Eq.~\eqref{eq:phidyn_c_tens}
by $Q$, then the {\em outer} equation at leading order will have the
same differential operator as the scalar model with constant $M$.

On the other hand, in the {\em inner} domain, we have, {\em provided}
$\Phinul$ solves the differential equation \eqref{eq:diffeq_Phi0},
$\epsilon^2(\nabla\Phi)^2 =\Phikrho^2 +O(\epsilon^2)=4
f(\Phi)+O(\epsilon)$ [this follows from Eqs.~\eqref{eq:nabphisq},
\eqref{eq:diffeq_Phi0_firstint}, and \eqref{eq:fphi}], whence
$Q\approx 1-\normhat:\normhat$.

%However, $Q$ is an approximation to $1-\normhat:\normhat$ that 
This approximation is accurate up to $O(\epsilon)$ only, which is not
sufficient, because the order $\epsilon$ correction would enter as a
bothersome additive term in the next-to leading order inner equation.

A better inner approximation to $1-\normhat:\normhat$ than just $Q$ is
provided by a minor modification.  Obviously, we have
$Q=1-\normhat:\normhat+O(\epsilon)\,\normhat:\normhat$ in the inner
region.  Taking this to some integer power $m$ we get, because
$1-\normhat:\normhat$ and $\normhat:\normhat$ are orthogonal
projectors:
\begin{equation}
\label{eq:Qm} 
Q^m =  1-\normhat:\normhat+O(\epsilon^m)\,\normhat:\normhat \>.
\end{equation}  

These considerations lead us to make the ansatz
\begin{eqnarray}
  \label{eq:phidyn_cval}
  \frac{\partial \phi}{\partial t} &=&
 M \nabla \cdot Q^m \nabla  \frac{1}{\epsilon^2}\,
\delta\mu(\nabla^2\phi,\phi) \>,\nonumber\\
\delta\mu(\nabla^2\phi,\phi) &\equiv&
%\bigg\{ 
-\epsilon^2\nabla^2 \phi
+ 2 f'(\phi)
 + \frac{\epsilon}{3\gamma} h'(\phi) \potelast%\bigg\}
\nonumber\\
\end{eqnarray}
%for the outer equations 
and leave the precise choice of the value of
$m$ for later -- it will be suggested by the asymptotic analysis.

The corresponding inner equations are
\begin{eqnarray}
  \label{eq:phidyn_c_innval}
  \partial_t \Phi -\frac{1}{\epsilon} v_n \partial_\rho \Phi &=&
M \nabla \cdot Q^m \nabla \frac{1}{\epsilon^2}\, 
\delta\mu(\nabla^2\Phi,\Phi)\>,\nonumber\\
 \delta\mu(\nabla^2\Phi,\Phi) &=&
%\bigg\{ 
-\frac{1}{\sqrt{g}}\,\partial_\rho \sqrt{g}\,
 \Phikrho \nonumber\\
 && - \epsilon^2\frac{1}{\sqrt{g}}\,\partial_\balpha
 \sqrt{g} g^{\balpha\bbeta}
 \Phikbbet \nonumber\\
&& + 2 f'(\Phi)
 + \frac{\epsilon}{3\gamma} h'(\Phi) \potelast%\bigg\}
\>,
\end{eqnarray}
with
\begin{align}
  \label{eq:diffop2_inn}
\nabla \cdot Q^m \nabla = &\frac{1}{\sqrt{g}} \partial_\nu \sqrt{g} g^{\nu\mu} 
\bigg\{\partial_\mu-\bigg[1-\Big(1-\frac{\epsilon^2 
(\nabla\Phi)^2}{4 f(\Phi)}\Big)^m\bigg] \nonumber\\
&\times\frac{1}{(\nabla\Phi)^2}\Phikmu g^{\alpha\beta} \Phikalph 
\partial_\beta\bigg\}
\>.
\end{align}

\setcounter{subsubsection}{0}
\subsection{Leading order}
In the outer equations, $Q$ becomes the identity operator to leading
order, i.e., $Q^{(0)}(\phinul) = 1$, and at the lowest order in
$\epsilon$, we have
\begin{equation}
  \label{eq:valout}
  \nabla^2 f'(\phinul) = 0 \>,
\end{equation}
a Laplace equation 
%for $f'(\phinul)$ 
that we know to be uniquely solvable for $f'(\phinul)$ with Dirichlet
boundary conditions at infinity.  This boundary condition is even
homogeneous (except possibly in a boundary region of order
$\epsilon$), leading to the unique solution $f'(\phinul)\equiv 0$.
This leaves us with the three possibilities $\phinul=0$,
$\nicefrac12$, $1$, of which $\phinul=0$ or $\phinul=1$ are realized,
according to the particular boundary condition on $\phinul$.

Again, $\phi=0$ and $\phi=1$ are solutions to the outer problem at all
orders of $\epsilon$.  Admittedly, the operator $Q$ becomes indefinite
at order $\epsilon^2$ for $\phi=0$ and $\phi=1$ [because of the
denominator $f(\phi)$], but this does not matter, since the
expression for $\delta \mu$ alone is zero already at $\phi=0$ and
$\phi=1$.

The leading-order inner equation reads [$g=1+O(\epsilon)$]
\begin{eqnarray}
  \label{eq:phydin_cval_innlead}
  \partial_\rho  \left[ 1-\frac{\left(\Phikrhonul\right)^2}{4 f(\Phinul)} 
\right]^m \!\! \partial_\rho 
\bigg(\partial_{\rho\rho} \Phinul -2 f'(\Phinul)\bigg) = 0\>.\nonumber\\
\end{eqnarray}
Clearly, this is solved by $\Phinul=\nicefrac12 \,(1-\tanh\rho)$,
which makes both the expression in brackets and in large parentheses
vanish.  If we require $m$ to be even, this solution is moreover
unique (up to translations, which are eliminated by the requirement
that the interface be at $\rho=0$).  For as soon as we assume
$(\Phikrhonul)^2 \ne 4 f(\Phinul)$, the $m$th power of the bracket
expression will be positive, allowing us to use similar arguments as
in Sec.~\ref{sec:expansions} between Eqs.~\eqref{eq:phydin_c_innlead}
and \eqref{eq:phydin_c_innlead_res} to prove that $\partial_{\rho\rho}
\Phinul -2 f'(\Phinul)=0$, and hence the bracket expression must be
zero, contrary to our assumption.  Thus we do get a definite solution
for $\Phinul$ from the inner equation, which moreover shows that at
leading order of the inner expansion the second-order $\rho$
derivatives of the operator $\nabla\cdot Q^m \nabla$ cancel each
other.
%  (This does not
% mean that the operator, considered as a functional of $\Phinul$ is
% identical to zero, only that it becomes zero for a particular argument
% function.)

\subsection{Next-to-leading order}

To simplify computations at the next order, we first expand
$\nabla\cdot Q^m\nabla$ up to formal order $\epsilon^0$.  This produces
\begin{align}
  \label{eq:expansion_nqmn}
\nabla&\cdot Q^m \nabla =  \nonumber\\
&\phantom{+}\frac{1}{\sqrt{g}}\,\partial_\bnu 
\sqrt{g} g^{\bnu\bmu} \partial_\bmu
+\frac{1}{\sqrt{g}}\, \frac{1}{\epsilon^2} \partial_\rho \sqrt{g}
\left(1-\frac{\epsilon^2 (\nabla\Phi)^2}{4 f(\Phi)}\right)^m \partial_\rho
\nonumber\\
&-\frac{1}{\sqrt{g}}\,\partial_\rho \sqrt{g} \frac{1}{\Phikrho} 
g^{\balpha\bbeta}\Phikbalph\partial_\bbeta
-\frac{1}{\sqrt{g}}\,\partial_\bnu\sqrt{g}  \frac{1}{\Phikrho}  g^{\bnu\bmu}
\Phikbmu \partial_\rho
\nonumber\\
&+\frac{1}{\sqrt{g}}\,\partial_\rho \sqrt{g}\frac{1}{\Phikrho^2}
\Phikbalph g^{\balpha\bbeta} \Phikbbet  \partial_\rho 
 + O(\epsilon)\>, 
\end{align}
Given that $\Phinul$ is a function of $\rho$ only, we realize that the
third and fourth terms on the right hand side are $O(\epsilon)$,
containing derivatives with respect to $\s$ or $\uc$ of $\Phi$, the
fifth is even $O(\epsilon^2)$, so these terms may be dropped
immediately in an expansion up to $O(1)$.  The second term on the
right hand side owes its existence the fact that $Q$ is not exactly the
projection operator on $\normhat$ [note that no such term is present
in Eq.~\eqref{eq:diffprolorder1}] and it has a prefactor of
$1/\epsilon^2$ due to the double derivative in $\rho$.  This term
which is desirable at leading order, because without it we would not
have a determinate zeroth order solution $\Phinul$, is somewhat
disturbing at the next order.  Since the order of this term is
$O(\epsilon^{m-2})$, we can make it small by choosing $m\ge 3$, i.e.,
restricting ourselves to even $m$ for the reasons discussed before, we
set $m=4$.  Then the only remaining term on the right hand side of
Eq.~\eqref{eq:expansion_nqmn} up to order $\epsilon^0$ is the first
term, which is the desired surface Laplacian.
% (the fifth term
% comes from the expansion of the denominator $(\nabla\Phi)^2$ and the
% $1/\epsilon^2$ has already been canceled by a prefactor $\epsilon^2$)

Using this result, we can write the next-to-leading (nontrivial) order
inner equation
\begin{eqnarray}
  \label{eq:nextorder}
 - v_n \partial_\rho \Phinul&=& M \frac{1}{\sqrt{g}}\,\partial_\bnu 
\sqrt{g} g^{\bnu\bmu} \partial_\bmu \delta\mu^{(1)}\>, 
\nonumber\\
\delta\mu^{(1)}&=&
%\left(\partial_{\s\s}+\partial_{\uc\uc} \right) 
-\bigg\{\linopl \Phione
 + (\kappa_1+\kappa_2) \partial_\rho \Phinul \nonumber\\
  &&\mbox{}\quad - \frac{1}{3\gamma} h'(\Phinul) 
\potelastnul\bigg\}
 \>,
\end{eqnarray}
again with $\linopl$ as given in Eq.~\eqref{eq:linop}.

Note that we actually seem to have skipped orders here.  The
leading-order inner equation is formally $O(\epsilon^{-4})$, but once
the zeroth-order inner solution is fixed, the differential operator
$\nabla\cdot Q^m \nabla$ is, according to \eqref{eq:expansion_nqmn},
of order $\epsilon^{\max(0,m-2)}$ only, so the order $\epsilon^{-3}$ vanishes
identically.  The order $\epsilon^{-2}$ is satisfied automatically,
because the zeroth-order chemical potential is zero; the next
nontrivial order is $\epsilon^{-1}$.  Alternatively, one may say that
the effective leading order has become $\epsilon^{-2}$.

Keeping only order one terms of the surface Laplacian, we have
$\nicefrac{1}{\sqrt{g}}\,\partial_\bnu \sqrt{g} g^{\bnu\bmu}
\partial_\bmu = \partial_{\s\s}+\partial_{\uc\uc}$, and the total
linear operator in front of $\Phione$ becomes
$-(\partial_{\s\s}+\partial_{\uc\uc})\linopl$.  It is hermitian,
because its hermitian factors commute.  Hence, $\partial_\rho\Phinul$
is a left null eigenfunction.  Multiplying \eqref{eq:nextorder} from
the left by it, integrating with respect to $\rho$ from $-\infty$ to
$\infty$, we obtain Eq.~\eqref{eq:veloc_scalmod}, with
$M=M^\ast\gamma$; to arrive at Eq.~(\ref{eq:vn_surfdiff}), we have to
set $M = D_s \gamma/\rho_s$. Together with App.~\ref{sec:sl_elastic},
this proves that the phase-field model of this section exhibiting a
modified tensorial mobility has the correct asymptotic behaviour for
small $\epsilon$, neither overconstraining the system by adding, nor
leaving it indeterminate by losing equations.

\section{Conclusions\label{sec:conclusions}}

The intuitive approach to constructing a phase-field model for surface
diffusion consists in using the chemical potential known from the
nonconservative model to define a current, involving its gradient and
a mobility that vanishes in the bulk phases, and to take the
divergence of this current as the time derivative of the phase field.
As has been shown in this article, this approach fails to produce the
correct asymptotics in a subtle way.  It does reproduce the equlibrium
limit correctly.

% Because the mobility remains a
% scalar in this approach, the corresponding model is designated as the
% scalar-mobility model.

Next, the idea was explored that this failure might be remedied by
making the mobility a tensor.  After all, surface diffusion may be
interpreted as highly anisotropic three-dimensional diffusion with a
diffusion tensor that has zero eigenvalue in one direction.  A
straightforward realization of this idea fails in a rather drastic
way, because restricting diffusion to the surfaces of constant phase
field does not impose any functional dependence of $\phi$ in the
normal direction.

A modification of the tensorial mobility however leads to a model
exhibiting the correct asymptotics.  This model is given by the
equations \eqref{eq:defQ}, \eqref{eq:phidyn_cval} with $m=4$,
\eqref{eq:defpotelast}, and \eqref{eq:elasteqs_ph}.  Larger even
values of $m$ also give valid models, a fact that might be useful in
optimizing numerical performance.  It is possible that $m=3$ will lead
to a valid model, too, but without positivity of the mobility tensor,
it is difficult to prove that the inner equation
\eqref{eq:phydin_cval_innlead} is solved by a unique phase-field
profile.

% Because the reason for failure is a subtle one

Since the condition \eqref{eq:veloceq0} destroying the asymptotic
correctness of the scalar-mobility model is a strong restriction, it
is conceivable that the model will satisfy the other equations while
not being asymptotic yet and therefore keep some of its utility.  It
should however be emphasized that no positive assertions concerning
its quantitative validity are available for this case.  The problem of
making reliable analytic statements about a nonasymptotic state is
rather difficult.

If the scalar-mobility model had a range of quantitative validity,
this would be restricted to some finite time interval, before the full
asymptotics kicks in.  Even then, it may be argued that the modified
tensorial model discussed here is probably more useful for numerical
simulations.  The point is the following.  In the scalar-mobility
model, the inner equation determining the interface velocity appears
only three orders in $\epsilon$ after the leading order.  A simulation
must of course represent the full model equations.  Suppose we wish
the interface velocity to be determined with an error not exceeding
order $\epsilon$.  Then the leading-order equation must be simulated
with an accuracy $O(\epsilon^4)$ at least.  If one takes a simple
second-order accurate discretization for the gradient operators in the
equations, the numerical error due to discretization alone will be
$O(a^2)$ for a grid spacing $a$.  To keep this smaller than
$\epsilon^4$, the grid spacing would have to scale with $\epsilon^2$,
which can be expected to lead to prohibitive computation times.
Therefore, in the scalar mobility model, high-accuracy discretizations
would be mandatory, even if the asymptotic error were controllable.

On the other hand, in the model discussed in Sec.~\ref{sec:correct},
as soon as the phase-field profile is represented with an error of
order $\epsilon$ or better, the effective leading order is only one
order lower than the one determining the interface velocity, similar
to the nonconservative case.  Hence, reasonable accuracy should be
attainable with grid spacings that scale as $\epsilon$, not as
$\epsilon^2$.  None of the benefits of high-accuracy discretizations
would be lost to the need of representing terms very accurately that
are very small in the leading-order equation.

Comparing the structure of the scalar-mobility model and the one based
on a modified tensorial mobility, it is clear that the former model
has much more intuitive appeal while the latter looks pretty
complicated.  But it has the virtue of producing the correct
sharp-interface asymptotics.  Very likely, in the future simpler
and more elegant models will turn up doing the same thing.  In the
meantime, this one may be of use.

% On the other hand, the model based modified tensorial mobility 

\textbf{Acknowledgments} Support of this work by DFG grants KA 672/5-4
and KA 672/9-1 is gratefully acknowledged.

\appendix

\section{Matching conditions\label{sec:matching}}

Let $\tilde\psi(x,y,z,t)=\psi(r,\s,\uc,t)$ be some arbitrary
(sufficiently often differentiable) function of space and time
obtained in solving the outer equations.  We write the corresponding
function of the inner solution as $\Psi(\rho,\s,\uc,t)$, and suppress
from now on, in this section, the dependence of functions on $\s$,
$\uc$, and $t$.  Moreover, we write the coefficient functions in
expansions with respect to $\epsilon$ with simple subscripts
indicating their order rather than superscripts in parentheses as in
the main text.  There, the notation is dictated by the fact that a
subscript would interfere with subscripts indicating tensor
components; here, a superscript would interfere with the primes
denoting derivatives.

We must have the asymptotic relationship
\begin{equation}
  \label{eq:asym_match1}
  \Psi(\rho) \sim \psi(r) = \psi(\epsilon\rho) 
\quad (\rho\to\infty,\,\epsilon\to0,\,\epsilon\rho\to0)\,.
\end{equation}
Expanding both functions in powers of $\epsilon$, we get
\begin{eqnarray}
  \Psi(\rho) &=& \Psi_0(\rho)+\epsilon\Psi_1(\rho)
                  +\epsilon^2\Psi_2(\rho)+\dots \>,    \label{eq:psi_exps1}\\
   \psi(\epsilon\rho) &=& \psi_0(r)+\epsilon\psi_1(r)
                  +\epsilon^2\psi_2(r)+\dots   \nonumber\\
               &=&  \psi_0(0)+\epsilon[\rho\psi'_0(0)+\psi_1(0)] \nonumber\\
                &&\mbox{}  +\epsilon^2[\rho^2\frac12\psi''_0(0)
         +\rho\psi'_1(0)+\psi_2(0)]+\dots \>, \nonumber\\
 \hspace*{1cm}\label{eq:psi_exps2}
\end{eqnarray}
where the derivatives are to be taken for $r\to +0$, should they be
discontinuous at $r=0$.  Analogous expressions with $r\to -0$ are
obtained for the asymptotics as $\rho\to-\infty$.

Equating powers of $\epsilon$, we then successively get the asymptotic
relationships
\begin{eqnarray}
 \lim_{\rho\to\pm\infty} \Psi_0(\rho) &=& \psi_0(\pm0)  \>, \label{eq:asymrels1}\\
 \Psi_1(\rho)  &\sim& \rho\psi'_0(\pm0)+\psi_1(\pm0) \quad (\rho\to\pm\infty)
\>,\nonumber\\ \label{eq:asymrels2}\\
 \Psi_2(\rho)  &\sim& \frac12\rho^2\psi''_0(\pm0)+\rho\psi'_1(\pm0)
  +\psi_2(\pm0) \nonumber\\
                  && \quad\quad\quad\quad\quad\quad    (\rho\to\pm\infty) \>.
\label{eq:asymrels3}
\end{eqnarray}
Moreover, asymptotic relations such as \eqref{eq:asymrels2} can be
decomposed into statements about function limits
\begin{eqnarray}
\lim_{\rho\to\pm\infty} \partial_\rho\Psi_1(\rho) &=& \psi'_0(\pm0) \>,
 \label{eq:asymrels2a} \\
\lim_{\rho\to\pm\infty} \left[\Psi_1(\rho)- \rho\psi'_0(\pm0)\right] 
 &=& \psi_1(\pm0) \>.
 \label{eq:asymrels2b}
\end{eqnarray}

\section{Local coordinates and metric\label{sec:loc_coord}}

Once the coordinates $\s$, $\uc$ on the interface have been chosen and
the third coordinate $r$ has been specified via \eqref{eq:posvec_loc},
the coordinate basis \eqref{eq:covar_bas} and, hence, the metric is
determined.  However, in order to obtain explicit expressions and
because it is desirable to interpret the metric in terms of familiar
geometric concepts, we need a few additional considerations.

For any curve $\rvec=\Rvec(s)$ on a surface, an accompanying
orthonormal frame can be defined consisting of the normal on the
surface $\norm$ at the considered curve point, its tangent
$\tang$, proportional to $\partial_s \Rvec(s)$ (and equal to it, if $s$
is the arclength), and the vector ${\mathbf l} = \norm\times\tang$.
This triple, in the older German literature \cite{dreszer75} sometimes
referred to as the Bonnet-Kowalewski trihedron, is more commonly known
under the name of Darboux frame \cite{spivak79}. It satisfies the
following equations:
\begin{equation}
\label{eq:darboux}
\left(\begin{array}{c}
  \partial_s \norm  \\
  \partial_s \tang \\
  \partial_s {\mathbf l} \\
\end{array}\right)
= \left(\begin{array}{ccc}
  0        & \kappa & \tau \\
  - \kappa & 0      & \alpha \\
  -\tau & -\alpha & 0 \\
\end{array}\right)
\left(\begin{array}{c}
  \norm  \\
  \tang \\
  {\mathbf l} \\
\end{array}\right) \>,
\end{equation}
where $\kappa$ is the normal curvature, i.e., the curvature of the
projection of the curve onto the plane spanned by its tangent and the
surface normal, $\alpha$ is the geodesic curvature, i.e., the
curvature of the projection of the curve onto the tangential plane of
the surface, and $\tau$ is the geodesic torsion.  The total curvature
$\kappa_s$ of the space curve given by $\Rvec(s)$ is related to the
normal and geodesic curvatures via
\begin{equation}
\label{eq:totcurv} \kappa_s^2 =\alpha^2+\kappa^2\>.
\end{equation}
In the mathematical literature \cite{dreszer75}, the sign conventions
for the normal curvature and geodesic torsion are opposite to the ones
used here.

If we identify our general curve on the surface with a coordinate curve
of constant $\uc$ from Sec.~\ref{sec:loccoord}, then the Darboux frame
will consist of the vectors $\norm$, $\tanga$, and $\tangb$ varying
with $\s$; if we identify it with a curve of constant $\s$, the Darboux
frame is given by $\norm$, $\tangb$, and $-\tanga$ (it is defined
right-handed).  This provides us with formulas for the derivatives of
the normal and tangent vectors of the surface,
\begin{align}
\frac{\partial \norm}{\partial \s}  & = \kappa_1 \tanga \> +\tau_1 \tangb \>, \label{eq:darboux1a}\\
\frac{\partial \tanga}{\partial \s} & = -\kappa_1 \norm \> +\alpha_1 \tangb \>,  \label{eq:darboux1b}\\
\frac{\partial \tangb}{\partial \s} &= -\tau_1 \norm \> - \alpha_1 \tanga \>, \label{eq:darboux1c} \\
\frac{\partial \norm}{\partial \uc} & = \kappa_2 \tangb \>- \tau_2 \tanga\>, \label{eq:darboux2a}\\
\frac{\partial \tangb}{\partial \uc}& =-\kappa_2 \norm \>- \alpha_2 \tanga \>,\label{eq:darboux2b}\\
\frac{\partial \tanga}{\partial \uc}&= \tau_2 \norm \> + \alpha_2 \tangb \>, \label{eq:darboux2c}
\end{align}
from which we immediately get the expressions (\ref{eq:covar_bas}) for
the vectors of the canonical coordinate basis $\{\Er,\Es,\Eu\}$.
Moreover, we obtain some simplifications.  Because $\Ealph =
\partial_\alpha \rvec$, we have the identities
\begin{equation}
\label{eq:mixedderiv}
\partial_\beta \Ealph = \partial_\alpha \Ebet \>.
\end{equation}
These are automatically fulfilled, if $\alpha=r$ or $\beta=r$, but
yield nontrivial relationships for $\alpha=\s$, $\beta=\uc$.

Consider a point $\rvec = \Rvec(\s,\uc)$ on the interface.  Then $\Es =
\partial_\s \Rvec(\s,\uc) = \tanga$, $\Eu= \partial_\uc \Rvec(\s,\uc)
=\tangb$, because we have specified our coordinates to be arclengths
(otherwise, $\Es$ and $\Eu$ would not be equal, just proportional to
the unit tangent vectors). From \eqref{eq:darboux1c} and
\eqref{eq:darboux2c}, we deduce that
\begin{equation}
\label{eq:esueqeus} \tau_2 \norm \> + \alpha_2 \tangb = -\tau_1 \norm
\> - \alpha_1 \tanga \>.
\end{equation}

Since the tangential vectors are linearly independent this implies
$\alpha_1=\alpha_2=0$, i.e., the geodesic curvature vanishes and our
coordinate curves are necessarily geodesics.  Then their normal
curvature becomes equal (up to a sign) to the spatial curvature, and
formulas \eqref{eq:darboux1a} through \eqref{eq:darboux2c} reduce to
the better known Frenet-Serret formulas \cite{dreszer75,spivak79}.
Secondly, we obtain from Eq.~\eqref{eq:esueqeus} that $\tau_1=-\tau_2$,
because the prefactors of the normal vector must match.

This means that the metric, Eq.~\eqref{eq:metric}, when restricted to
the interface, will depend on just three parameters, viz. $\kappa_1$,
$\kappa_2$, and $\tau$.  But this is precisely the number of
independent elements of a symmetric $2\times2$ tensor, so we can
describe the most general metric this way locally.

To complete the discussion of geometric relationships near the
interface, let us note that  formula \eqref{eq:divergence} applied to
$\Avec=\Er$ at $r=0$, yields
\begin{equation}
\kappa \equiv \nabla\cdot\norm = \kappa_1+\kappa_2 \>,
\end{equation}
i.e., the mean curvature $\kappa$ is the sum of {\em any} two normal
curvatures corresponding to orthogonal directions, not just the sum of
the two principal curvatures.  This is the ultimate justification for
using the same notation for the principal curvatures in
Sec.~\ref{sec:sharp_i_mod} and the normal curvatures in
Sec.~\ref{sec:loccoord}.

Another interesting geometric quantity is the Gaussian curvature
$\kappa_G$, given as the product of the two {\em principal} curvatures.
A general formula in terms of the Darboux frame is
\begin{align}
\label{eq:gauss_curv} \kappa_G &= \det \left(\norm, (\tanga\cdot\nabla)
\norm, (\tangb\cdot\nabla) \norm \right) = \norm \left(\partial_\s
\norm \times \partial_\uc \norm \right) \>.
\end{align}
Evaluating this with the help of \eqref{eq:darboux1a} and
\eqref{eq:darboux2a}, we find an expression for the Gaussian curvature
in terms of normal curvatures and torsions %of the orthogonal geodesics
\begin{align}
\label{eq:gauss_curv_eval}
 \kappa_G = \kappa_1 \kappa_2 - \tau^2 \>.
\end{align}
From this, one may conclude that the torsions have to vanish when
$\kappa_1$ and $\kappa_2$ become the principal curvatures, i.e., take
their extremal values.

\section{Sharp-interface limit of the elastic part of the model\label{sec:sl_elastic}}

Loosely speaking, the title of this appendix refers to two things.  On
the one hand, it must be shown that Eq.~\eqref{eq:elasteqs_ph} reproduces
the bulk elastic equations as well as the boundary conditions for the
elastic problem.  On the other hand, $\potelastnulbar$ must be
evaluated and shown to be the elastic contribution to the chemical
potential of Eq.~\eqref{eq:delta_mu}.

The first task is easily accomplished.  Inserting the
outer solutions $\phi=1$ and $\phi=0$ into \eqref{eq:elasteqs_ph}, we
obtain
\begin{eqnarray}
  \label{eq:elast_sh1}
  \sum_j \frac{\partial\sigma_{ij}}{\partial x_j} &=& 0 \>, \\
  \label{eq:elast_sh2}
  \frac{\partial p}{\partial x_i} &=& 0 \>, 
\end{eqnarray}
which are the mechanical equilibrium conditions in the solid and the
liquid, respectively.  The boundary conditions follow from the inner
equation \eqref{eq:elasteq_inn}, which at leading order reduces to
\begin{eqnarray}
  \label{eq:elasteq_innlead}
  0 &=& \partial_\rho  
     \left(\Ssigma_{\gamma r}^{(0)} \Ehgam\right)
 + O(\epsilon)  \nonumber\\
 &=&  \norm \partial_\rho \Ssigma_{r r}^{(0)} 
+\tanga \partial_\rho \Ssigma_{\s r}^{(0)} 
+ \tangb\partial_\rho \Ssigma_{\uc r}^{(0)} + O(\epsilon) \>,
\end{eqnarray}
where we have used that the $\Ehgam$ become the basis vectors of
the Darboux frame at leading order and that these are independent of
$r$ and, hence, $\rho$.  From \eqref{eq:elasteq_innlead}, we have
\begin{equation}
  \label{eq:zerosigma}
   \partial_\rho \Ssigma_{r r}^{(0)} = \partial_\rho \Ssigma_{\s r}^{(0)}
= \partial_\rho \Ssigma_{\uc r}^{(0)} = 0 \>,
\end{equation}
i.e., the three appearing components of the generalized stress tensor
are independent of $\rho$.  The matching conditions then provide us
with
\begin{eqnarray}
  \label{eq:lim_bndcond}
  \sigma_{nn} &=& \lim_{\rho\to-\infty}\Ssigma_{r r}^{(0)} 
%= \Ssigma_{r r}^{(0)} (\s,\uc)
 =  \lim_{\rho\to\infty}\Ssigma_{r r}^{(0)} 
= -p \>, \\
 \label{eq:lim_bndcond1}
 \sigma_{nt_1} &=&  \lim_{\rho\to-\infty}\Ssigma_{r \s}^{(0)} 
 =  \lim_{\rho\to\infty}\Ssigma_{r \s}^{(0)} = 0  \>,   \\
 \label{eq:lim_bndcond2}
 \sigma_{nt_2} &=& \lim_{\rho\to-\infty}\Ssigma_{r \uc}^{(0)} 
 =  \lim_{\rho\to\infty}\Ssigma_{r \uc}^{(0)} = 0     \>, 
\end{eqnarray}
with the limit values in the nonsolid phase ($\rho\to\infty$)
following directly from \eqref{eq:elasteqs_ph}.  At the next order,
the phase-field equations yield the capillary overpressure correction
to \eqref{eq:lim_bndcond}, but we shall not consider this level of
detail here.

For the second task, we need the strain tensor ${\mathcal U}$ in curvilinear
coordinates to evaluate \eqref{eq:defpotelast} in the inner domain.
To perform the transformation, we first note that ($\Ealph^i$ is the
$i$th cartesian component of $\Ealph$)
\begin{eqnarray}
  \label{eq:strain_curv_def}
  U_{\alpha\beta} &=& \Ealph {\mathcal U} \Ebet = U_{ij}  \Ealph^i \Ebet^j \nonumber\\
&=& \frac12 \left(U_{i,j}+U_{j,i}\right) 
\frac{\partial x^i}{\partial\alpha}  \frac{\partial x^j}{\partial\beta} 
\nonumber\\
&=& \frac12\left( \frac{\partial U_i}{\partial\beta} 
\frac{\partial x^i}{\partial\alpha} 
+\frac{\partial U_j}{\partial\alpha}  \frac{\partial x^j}{\partial\beta} 
\right) \nonumber\\
&=&   \frac12\left[ \partial_\beta  \big(U_i \Ealph^i\big) 
-  U_i \partial_\beta \Ealph^i + \partial_\alpha \big(U_j \Ebet^j\big) 
-  U_j \partial_\alpha \Ebet^j
\right]  \nonumber\\
&=&   \frac12\left[ \partial_\beta  U_\alpha 
-  \Uvec \partial_\beta \Ealph + \partial_\alpha  U_\beta 
-  \Uvec \partial_\alpha \Ebet 
\right]\>, 
\end{eqnarray}
where some additional notations should be obvious: $U_i$ is a cartesian
component of the displacement vector, $\Uvec$ is the full vector.
Introducing the connection coefficients or Christoffel symbols (of the
second kind)
\begin{equation}
  \label{eq:christoffel}
  \Gammagab = \Ehgam \partial_\beta \Ealph \>,
\end{equation}
we can rewrite \eqref{eq:strain_curv_def} as
\begin{eqnarray}
    \label{eq:strain_curv_cov}
     U_{\alpha\beta} &=&  
  \frac12 \left(U_{\alpha,\beta}+U_{\beta,\alpha}\right)
 -\frac12 \Gammagab U_\gamma -\frac12 \Gammagba U_\gamma \nonumber\\
  &=& \frac12 \left(U_{\alpha;\beta}+U_{\beta;\alpha}\right) \>,
\end{eqnarray}
where the semicolon in the last line denotes a covariant derivative.
This last result is evident - tensorial equations retain their
invariant form when written with covariant derivatives.  Since in a
flat space the derivative with respect to a cartesian coordinate is
automatically a covariant derivative, we could have written down the
last equation from knowledge of the form of the small-strain tensor in
cartesian coordinates.

The fastest way to calculate the Christoffel symbols seems to be via
their definition \eqref{eq:christoffel}.  There is an alternative
formula using derivatives of the metric coefficients and known to most
physicists from general relativity courses, but while this produces
the same results, it does so via lengthy intermediate expressions
only.  Since we will consider only the leading-order expression for
$\potelast$, we give the connection coefficients only to that order as
well.  At order $\epsilon^0$, the metric \eqref{eq:metric} becomes the
unit tensor, so superficial consideration might suggest that, being
defined via derivatives of the metric, the Christoffel symbols should
all vanish.  However, one has to be careful to first take derivatives
and then collect terms of order one, because derivatives with respect
to $r$ produce a factor $1/\epsilon$ when rewritten in terms of the
stretched coordinate $\rho$.  Taking this into account, we find the
following Christoffel symbols to generically be nonzero at leading
order:
\begin{eqnarray}
  \label{eq:christoffel_0}
  \Gammarss &=& -\kappa_1 \>, \nonumber\\
  \Gammaruu &=& -\kappa_2 \>, \nonumber\\
  \Gammarsu &=& \Gammarus = -\tau \>, \nonumber\\ 
  \Gammasrs &=& \Gammassr = \kappa_1 \>, \nonumber\\
  \Gammasru &=& \Gammasur = \tau \>, \nonumber\\ 
  \Gammaurs &=& \Gammausr = \tau \>, \nonumber\\
  \Gammauru &=& \Gammauur = \kappa_2 \>.
\end{eqnarray}
All the remainig ones are either $O(\epsilon)$ or exactly zero.

The components of the strain tensor are
\begin{eqnarray}
  \label{eq:strain_curv_0}
  \Urr &=& \frac{1}{\epsilon} U_{r,\rho}\>, \nonumber\\
  \Uss &=& U_{\s,\s} + \kappa_1 U_r +O(\epsilon) \>, \nonumber\\
  \Uuu &=& U_{\uc,\uc} + \kappa_2 U_r +O(\epsilon) \>,  \nonumber\\
  \Urs &=& \frac12 \left(U_{r,\s}
       +\frac{1}{\epsilon}U_{\s,\rho}\right) 
    - \kappa_1 U_\s - \tau U_\uc   +O(\epsilon)  \>, \nonumber\\
  \Uru &=& \frac12 \left(U_{r,\uc}
       +\frac{1}{\epsilon}U_{\uc,\rho}\right) 
    - \kappa_2 U_\uc - \tau U_\s   +O(\epsilon)  \>, \nonumber\\
  \Usu &=& \frac12 \left(U_{\s,\uc}+U_{\uc,\s}\right) 
    + \tau U_r    +O(\epsilon) \>.
\end{eqnarray}
Since the left hand sides of these equations are of order 1, the terms
having a factor $1/\epsilon$ can be of order 1 at most, too.  This
implies $\Unul_{r,\rho}=\Unul_{\s,\rho}=\Unul_{\uc,\rho} = 0$, hence
the zeroth-order displacement components must not depend on $\rho$:
\begin{eqnarray}
  \label{eq:displ_0}
   \Unul_r &=& \Unul_r(\s,\uc) \>, \nonumber\\
   \Unul_\s &=& \Unul_\s(\s,\uc) \>, \nonumber\\
   \Unul_\uc &=& \Unul_\uc(\s,\uc) \>.
\end{eqnarray}
From this and \eqref{eq:strain_curv_0}, we can immediately conclude
that several of the strain tensor components are independent of $\rho$
at zeroth order as well:
\begin{eqnarray}
  \label{eq:strain_0sp}
   \Ussnul &=& \Ussnul(\s,\uc) \>, \nonumber\\
   \Uuunul &=& \Uuunul(\s,\uc) \>, \nonumber\\
   \Usunul &=& \Usunul(\s,\uc) \>.
\end{eqnarray}
The new feature in comparison with the two-dimensional case is that
even a shear component of the strain tensor will be continuous across
the interface, i.e., in a simulation the nonsolid phase will possibly
undergo a biaxial strain.  This is not really a problem, keeping in
mind that this phase is treated in simulations as if it were a solid
phase with vanishing shear modulus.  Shear strains in this phase do
not correspond to anything physical, and there is neither a shear
stress nor energy associated with them.

If we take $\lim_{\abs{\rvec}\to\infty} p = p_0$ as boundary condition
for the pressure in the nonsolid bulk phase, then we know from
Eq.~\eqref{eq:elast_sh2} that the pressure will take that value
everywhere, and we can integrate the equation for $\Ssigma_{r r}$ from
\eqref{eq:zerosigma} with a known integration constant.  Writing the
result out in terms of the strains [using \eqref{eq:hooke}], we
obtain, introducing the abbreviation $\hnul = h(\Phinul)$:
\begin{align}
  \label{eq:sigmarr0}
 \hnul &\left[2\mus \Urrnul
 + \lambda\left(\Urrnul+\Ussnul+\Uuunul\right)\right] \nonumber\\
 & + (1-\hnul)\lamtil\left(\Urrnul+\Ussnul+\Uuunul\right) -p_0 = -p_0 \>.
 \nonumber\\
\end{align}
We know the $\rho$ dependence of all terms in this equation except
$\Urrnul$, which suggests to solve for the latter quantity.  This
results in
\begin{equation}
  \label{eq:urr0}
  \Urrnul = - \left(\Ussnul+\Uuunul\right)\left(1-
   \frac{2\mus\hnul}{(2\mus+\lambda-\lamtil)\hnul+\lamtil}\right) \>.
\end{equation}
The equations for $\Ssigma_{\s r}^{(0)}$ and $\Ssigma_{\uc r}^{(0)}$
from \eqref{eq:zerosigma} imply together with the boundary conditions
\eqref{eq:lim_bndcond1} and \eqref{eq:lim_bndcond2} that $\Unul_{\s r}
= \Unul_{\uc r} =0$.  Then Eq.~\eqref{eq:defpotelast} turns into
\begin{align}
  \label{eq:potelast0}
 \potelastnul  = & \>\mus \left[\left(\Urrnul\right)^2+\left(\Ussnul\right)^2
+\left(\Uuunul\right)^2 + 2 \left(\Usunul\right)^2
\right] \nonumber\\
& + \frac{\lambda-\lamtil}{2} \left(\Urrnul+\Ussnul+\Uuunul\right)^2 \>,
\end{align}
and we can conclude from \eqref{eq:urr0} and \eqref{eq:strain_0sp}
that the elastic potential depends on $\rho$ only via $h_0$.
Abbreviating $tr_{\rm 2D} \Unul = \Ussnul+\Uuunul$ and $X =
(2\mus+\lambda-\lamtil)\hnul+\lamtil$, we get the compact form
\begin{align}
  \label{eq:potelast0a}
 \potelastnul  = & \mus \left(tr_{\rm 2D} \Unul\right)^2 
\left[1-\frac{2\mus \hnul}{X}-\frac{2\mus \hnul\lamtil}{X^2}\right]
 \nonumber\\ 
&+\mus \left[\left(\Ussnul\right)^2+\left(\Uuunul\right)^2 
+ 2 \left(\Usunul\right)^2\right]\>.
\end{align}
According to \eqref{eq:typicalints} and \eqref{eq:intdphirho2}, we
have 
\begin{equation}
  \label{eq:potelast_reduce_h}
\potelastnulbar =\frac{\int_{-\infty}^{\infty}\potelastnul
\left(\partial_\rho \Phinul\right)^2 d\rho}{\int_{-\infty}^{\infty}
\left(\partial_\rho \Phinul\right)^2 d\rho} = \int_0^1 \potelastnul d\hnul \>,
\end{equation}
so the only integral that actually necessitates a calculation when
integrating \eqref{eq:potelast0a} is
\begin{align}
  \label{eq:integ_potel}
 \int_0^1 &\left(\frac{2\mus \hnul}{X}
  +\frac{2\mus \hnul\lamtil}{X^2} \right)d\hnul \nonumber\\
&= \frac{2\mus}{(2\mus+\lambda-\lamtil)^2}\int_{\lamtil}^{2\mus+\lambda} 
\frac{(X-\lamtil)(X+\lamtil)}{X^2} dX  \nonumber\\
&= \frac{2\mus}{2\mus+\lambda} \>,
\end{align}
all the other integrals reduce to integrating a constant from zero to one.

We obtain
\begin{align}
  \label{eq:potelast_barval}
  \potelastnulbar = \frac{\mus\lambda}{2\mus+\lambda} \left(tr_{\rm
      2D} \Unul\right)^2 + \mus \sum_{\balpha,\bbeta} 
     \left(\Unul_{\balpha\bbeta}\right)^2 \>. 
\end{align}
Note that both terms are interface scalar invariants, taking the same
form in any (2D) coordinate system.

To relate this to the sharp-interface formula \eqref{eq:delta_mu}, we
should transform it into an expression in terms of surface quantities,
belonging to the outer equations. We have
\begin{eqnarray}
  \label{eq:outerform}
  \ussnul &=& \Ussnul(\s,\uc) \>, \nonumber\\
  \uuunul &=& \Uuunul(\s,\uc) \>, \nonumber\\
  \urrnul &=& -\frac{\lambda}{2\mus+\lambda} \left(\Ussnul+\Uuunul\right) \>,
\end{eqnarray}
the last equation following from \eqref{eq:urr0} by taking the limit
$\rho\to-\infty$.

In the following, we drop the superscript $(0)$. Hooke's law together
with \eqref{eq:outerform} provides us with
\begin{eqnarray}
  \label{eq:stressform}
  \uss+\uuu &=& \frac{1-\nu}{E} \left(\sigss+\siguu-2\sigrr\right) \>,
\nonumber\\
  \uss-\uuu  &=& \frac{1+\nu}{E} \left(\sigss-\siguu\right) \>,
\end{eqnarray}
and the prefactors are related to the Lamé constants by
%$\lambda/(2\mus+\lambda) = \nu/(1-\nu)$, 
$(1-\nu)/E = (2\mus+\lambda)/[2\mus(2\mus+3\lambda)]$ and $(1+\nu)/E =
1/2\mus$.

Inserting \eqref{eq:stressform} into \eqref{eq:potelast_barval} and
renaming $\sigrr$ into $\signn$, it is straightforward to show that
\begin{align}
  \label{eq:potelast_barsh}
  \potelastnulbar =& \frac{1+\nu}{2E} \sum_{\balpha,\bbeta} 
 \left(\sigbalphbbet - \signn \delta_{\balpha\bbeta}\right)^2 \nonumber\\ 
 & - \frac{\nu}{2E} \Big[\sum_\balpha \big(\sigbalphbalph-\signn\big)\Big]^2
\>,
\end{align}
which is, up to a factor, of $1/\rho_s$ the elastic part of the
chemical potential of Eq.~\eqref{eq:delta_mu}, just written in
curvilinear coordinates instead of cartesian ones.  It is clear from
this result that if we set [in Eq.~\eqref{eq:phidyn_cval}] $M=\gamma
M^\ast = \gamma D_s/\rho_s$, Eq.~\eqref{eq:veloc_scalmod} will become
identical to Eq.~\eqref{eq:vn_surfdiff} from the sharp-interface
model, which completes the proof that the phase-field model of
Sec.~\ref{sec:correct} asymptotically approaches the sharp-interface
model of Sec.~\ref{sec:sharp_i_mod}.

\section{Useful properties of the phase field functions\label{sec:collection}}

In order to simplify it for the reader to find the actual
relationships for the various functions involving the phase-field that
are used in the text, they are collected here for reference (and
concreteness). Often only certain properties but not the precise form
of these functions is important.

The chosen double-well potential is 
\begin{equation}\label{eq:fphi}
    f(\phi) = \phi^2 (1-\phi)^2 \>.
\end{equation}
Its derivative is given by
\begin{equation}\label{eq:fprimephi}
    f'(\phi) = 2 \phi (1-\phi) (1-2\phi) \>.
\end{equation}
The ``volume fraction function'' $h(\phi)$ is 
\begin{equation}\label{eq:hphi}
    h(\phi) = \phi^2 (3-2\phi) \>,
\end{equation}
having the derivative
\begin{equation}\label{eq:hprimephi}
    h'(\phi) = 6 \phi (1-\phi)  \>.
\end{equation}
Then a useful observation is that
\begin{equation}\label{eq:fphih}
    f(\phi) = \left(\frac16 h'(\phi)\right)^2.
\end{equation}

To solve the ordinary differential equation satisfied by the
zeroth-order inner solution
\begin{equation}\label{eq:diffeq_Phi0}
    \partial_{\rho\rho} \Phinul -2 f'(\Phinul) = 0 \>,
\end{equation}
with boundary conditions $\lim_{\rho\to-\infty}\Phinul(\rho) = 1$ and
$\lim_{\rho\to\infty}\Phinul(\rho) = 0$, we multiply by $\partial_\rho
\Phinul$, integrate and take the square root (with the correct sign) to
obtain
\begin{equation}\label{eq:diffeq_Phi0_firstint}
    \partial_\rho \Phinul = -2 \Phinul (1-\Phinul) 
= -\frac13 h'(\Phinul)
    \>,
\end{equation}
which can be solved by separation of variables. The solution is, up to
a translation in $\rho$, given by
\begin{equation}\label{eq:solution_Phi0}
    \Phinul = \frac12 (1-\tanh\rho) \>.
\end{equation}
Requiring the position of the interface to be at $\rho=0$ fixes the
choice out of the one-parameter set of solutions, present due to the
translational invariance of the differential equation.

With the help of the second equality of
\eqref{eq:diffeq_Phi0_firstint}, it is easy to calculate certain
integrals appearing in the asymptotic analysis.  Those integrals
typically contain the factor $\left(\partial_\rho \Phinul\right)^2$;
to do the integral, it is then beneficial to replace one of the
factors (and only one) with $-h'(\Phinul)/3$.  Integrals obtained this
way have the structure
\begin{align}
 \label{eq:typicalints}
I = &\int_{-\infty}^{\infty}\,d\rho\> f(h(\Phinul))\, 
\,\left(\partial_\rho \Phinul\right)^2 
 \nonumber\\
 = & -\frac13 \int_{-\infty}^{\infty}\,d\rho\> f(h(\Phinul))\, h'(\Phinul) 
\,\partial_\rho \Phinul
\nonumber\\
 = & -\frac13\int_1^0 \,d\Phinul \>  f(h(\Phinul)) h'(\Phinul) 
= \frac13\int_0^1 
\,d\hnul\> f(\hnul)
\>.
\end{align}
This way, one arrives, for example, at
\begin{equation}
  \label{eq:intdphirho2}
  \int_{-\infty}^{\infty} \left(\partial_\rho \Phinul\right)^2
 d\rho = \frac13 \>.
\end{equation}

% Integrals that appear in 
% \begin{eqnarray}
% I_1 &=&  -\int_{-\infty}^{\infty}\,d\rho\>  h'_0  \partial_\rho \Phi_0 = 1 \>,
% \label{I1} \\
% I_2 &=& -\int_{-\infty}^{\infty}\,d\rho\> U_{\rho\rho}^{(0)}  h'_0  
% \partial_\rho \Phi_0 \>, \label{I2} \\
% I_3 &=& -\int_{-\infty}^{\infty}\,d\rho\> {U_{\rho\rho}^{(0)}}^2  h'_0  
% \partial_\rho \Phi_0 \>. \label{I3}
% \end{eqnarray}

\end{document}